\documentclass[a4paper,11pt]{article}
\pdfoutput=1 

\usepackage{jinstpub} 
\usepackage{lineno,hyperref}
\modulolinenumbers[1]
\usepackage{graphicx}
\usepackage{float}
\usepackage{threeparttable}
\usepackage{subfigure}
\usepackage{amsmath,amsfonts,amsthm} 
\usepackage[version=4]{mhchem}
\usepackage[load-configurations=abbreviations,detect-all=true]{siunitx}

\title{\boldmath Three-dimensional convolutional neural networks for neutrinoless double-beta decay signal/background discrimination in high-pressure gaseous Time Projection Chamber}

\author[a]{Pengcheng Ai,}
\author[a,1]{Dong Wang,\note{Corresponding author.}}
\author[a]{Guangming Huang}
\author[a]{and Xiangming Sun}

\affiliation[a]{Central China Normal University, No.152 Luoyu Road, Wuhan, Hubei, 430079 P.R.China}

\emailAdd{dongwang@mail.ccnu.edu.cn}

\abstract{In the search for neutrinoless double-beta decay, the high-pressure gaseous Time Projection Chamber has a distinct advantage, because the ionization charge tracks produced by particle interactions are extended and the detector captures the full three-dimensional charge distribution with appropriate charge readout systems.  Such information of tracks provides a crucial extra-handle for discriminating signal events against backgrounds.  In this paper, we constructed a toy model to demonstrate where the discrimination power comes from and how much of it the neural network models have already harnessed.  Then we adapted 3-dimensional convolutional and residual neural networks on the simulated double-beta and background charge tracks and tested their capabilities in classifying these two types of events.  We show that both the 3D structure and the overall depth of the neural networks significantly improve the accuracy of the classifier and lead to results better than previous works.  We also studied their performance under various spatial granularities as well as different diffusion and noise conditions.  The results indicate that the methods are stable and generalize well despite varying experimental conditions.}


\arxivnumber{1803.01482} 

\begin{document}
\maketitle
\flushbottom

\section{Introduction}

In particle physics, the Standard Model deals with the basic elements of matter, fundamental interactions and the mechanism responsible for the origin of mass. The Standard Model is so successful that the recent confirmation of the Higgs boson at the Large Hadron Collider gave credit to the theoretical prediction which had been proposed over fifty years ago. However, some basic problems are left unsolved, especially regarding the asymmetry of matter and antimatter in the universe. Beyond the Standard Model, A new kind of mass mechanism, called the Majorana mass mechanism, provides a latent solution for the matter-antimatter asymmetry in the early universe by introducing the leptogenesis process. A feasible method to testify the mechanism relies on the experiment of neutrinoless double-beta decay ($0\nu\beta\beta$), which demonstrates that the neutrino is its own antiparticle \cite{schechter1982neutrinoless}, the core inference of Majorana's theory.

Although the theory itself is self-consistent, the former experiments \cite{gando2016search,andringa2016current,abgrall2014majorana,agostini2016search,artusa2015searching,pocar2015searching} in the last century and the beginning of this century did not give a statistically significant result, which is generally attributed to the rarity of such events (or equivalently, the long lifetime). Among the experiments currently done and planned in the future, the high-pressure gaseous Time Projection Chamber (TPC) experiments \cite{gomez2014present,chen2017pandax} which feature mass scale-up and high energy resolution have the best potential to fulfill all the requirements with positive discovery. Besides, the \ce{^{136}Xe} isotope in its gaseous form is a good candidate to act as the working gas for such experiments. Since the square of sensitivity is inversely proportional to the background rate in the presence of non-zero backgrounds \cite{alessandria2013sensitivity}, the background rejection in the TPC experiments becomes a central problem when the experimental settings are given.

A "signal event" in the high-pressure xenon gas is the double-beta decay event. Electrons moving through the xenon lose energy moderately and leave a track of ionization at an approximately fixed rate until they reach the end of the track. As they approach the end, a significant amount of energy is deposited and the ionization density rises sharply before they become completely non-relativistic. This phenomenon, nominally called the Bragg peak, is an effective method to determine the presence of an electron emitted by the double-beta decay of the \ce{^{136}Xe} isotope. In contrast to the signal event, which leaves two "blobs" due to high ionization density at the ends of the tracks of two electrons, a "background event" is usually an event with similar total energy but one "blob", for only one electron is involved in the physical process. The background events are mainly generated from the radioactive impurities in the detector materials and radiation originating from outside the instrument. The dominant background comes from the photoelectric interaction or the Compton effect of high energy gammas emitted by \ce{^{214}Bi} or \ce{^{208}Tl} isotopes. Due to multiple Coulomb scattering, the tracks of these signal and background events are not straight in the dense gas; however, the patterns of events are distinguishable from a topological view.

The utilization of these peculiar signatures to improve the background rejection performance was discussed and studied in previous TPC experiments \cite{luescher1998search,ferrario2016first,martin2016sensitivity}. Besides, some recent studies demonstrate that convolutional neural networks have been applied successfully to real events in TPCs \cite{Acciarri:2016ryt}. However, to the best of our knowledge, no theoretical analysis of the discrimination power of the neural network model is discussed in previous studies. In this paper, we want to address both the theoretical question and the practical issues about applying the neural network model to TPC experiments. Section \ref{section:Instrument conditions} briefly introduces the conceived instrument used for this experiment. Section \ref{section:Deep learning method} explains the deep learning method we use in detail. Section \ref{section:Network architectures} lists all the network architectures used in the paper. Section \ref{section:A toy model demonstration} discusses how to use an abstract toy model to analyze the theoretical limit of the probabilistic model and the capability of the neural network to utilize the presented information. Section \ref{section:Data simulation and preprocessing} gives the details of generating and preprocessing the data. Section \ref{section:Simulation studies} discusses the simulations we conduct in different settings based on the previous data and method. Finally, a conclusion is drawn in Section \ref{section:Discussion and conclusion}.

\section{Instrument conditions}
\label{section:Instrument conditions}

\begin{figure}[htbp]
	\centering	
	\includegraphics[width=0.9\textwidth]{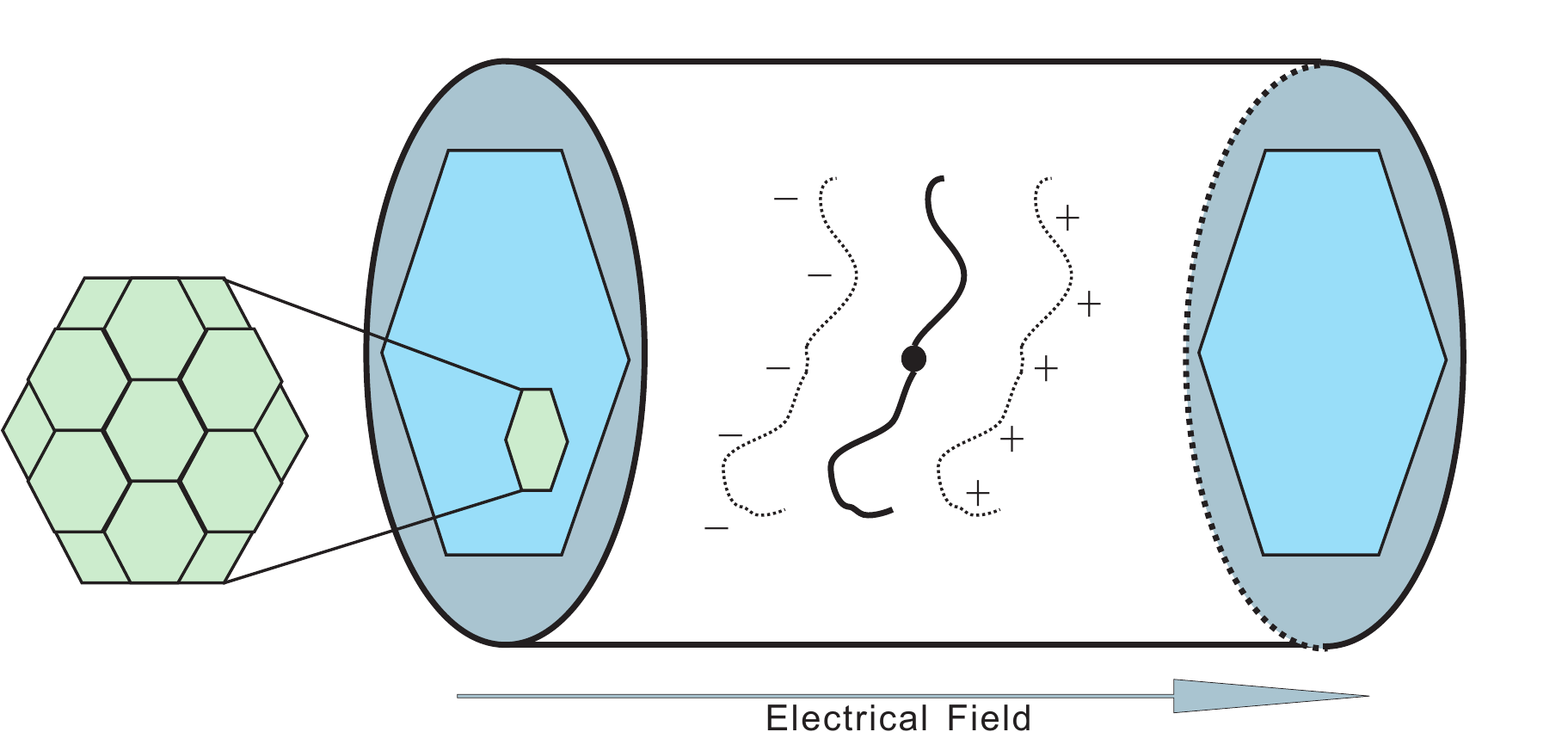}		
	\caption{\label{fig:TPC} The conceptual design of a TPC with the \emph{Topmetal} charge readout plane.}
\end{figure}

The Time Projection Chamber (TPC) is an instrument which uses a sensitive volume of medium (gas, liquid) to perform 3D reconstruction of particle tracks or interactions, in addition to some kinetic observables, if electric fields and/or magnetic fields apply. When passing through the medium, a particle will produce the primary ionization along its track. Under the electric field ($\sim500 V/cm$) in the active region, charges will drift to two side planes of the chamber. The z coordinate (longitudinal) is measured by the drifting time, calculated as the difference between the arrival time and $t_0$ from an independent detector. The information of $t_0$ can also come from kinetic constraints (for example, the time it takes for charges to reach the anode side plane and the cathode side plane should satisfy a certain relationship). The (x, y) coordinates (transverse) are directly measured by the readout plane. The shape of the chamber could be cylindrical, which is most commonly used and able to maximize the untouched events (tracks unfettered by the edges of the chamber) with a limited volume of medium. In the case of $0\nu\beta\beta$, the medium \ce{^{136}Xe} serves as both the producer of the energetic charged particle and the provider of the ionization charges.

A conceptual model of a TPC for $0\nu\beta\beta$ experiments is shown in Fig. \ref{fig:TPC}. In this model, we use an advanced technology called \emph{Topmetal} \cite{an2016low,zou2016test} to build charge readout planes arranged in the hexagonal pattern. The \emph{Topmetal} chip can collect charges directly without avalanche multiplication, which provides an excellent Equivalent Noise Charge performance (less than $15e^{-}$). Besides, the pixelated charge readout plane makes it possible to simultaneously achieve sufficient energy resolution and spatial resolution for tracking.

To implement such an instrument, several aspects should be taken into consideration:

\begin{itemize}
	\item \emph{Isotope medium}. For \ce{^{136}Xe} at 10 bar, both transverse diffusion and longitudinal diffusion (see Section \ref{subsection:Diffusion and noise simulation}) can be significant. This will blur the actual track of electrons emitted by the signal event.
	
	\item \emph{Charge Sensitive pre-Amplifier (CSA) of the Topmetal chip}. It is possible to design a CSA with noise of tens of electrons using the \emph{Topmetal} technology. However, in the detailed analysis, its impact is quantifiable and should not be neglected (see Section \ref{subsection:Diffusion and noise simulation}).
	
	\item \emph{Pixel pitch}. The pixel pitch is relevant to the background discrimination efficiency (see Section \ref{subsection:Granularity simulation}) we can achieve. Generally speaking, using a smaller pixel pitch is beneficial to background rejection by the neural network model. However, it will suffer from the effect of diffusion and noise more seriously. The optimal value is a result of the trade-off among the former two aspects and the effect of background rejection.
\end{itemize}

Physical characteristics of the $0\nu\beta\beta$ events set the basic specifications of the instrument. For example, the intrinsic energy resolution for 10 bar \ce{^{136}Xe} at the $Q_{\beta\beta}$ is 0.3\% Full Width at Half Maximum (FWHM) \cite{alvarez2013near}, which corresponds to a total fluctuation of about 130 $e^-$. In order to achieve 1\% FWHM resolution, the noise allowed for all the pixels in the readout plane is about 400 $e^-$. If 150 pixels are active in a $0\nu\beta\beta$ event, the maximum allowed per pixel noise (along the z axis of the TPC) is about 30 $e^-$. This motivates us to choose 30 $e^-$ as the worst noise level in the simulation study. In contrast to noise, diffusion is harder to quantify because of its close relationship to the implementation of the instrument. For 1m drift in pure xenon, the transverse diffusion in a 500 V/cm drift field is $\sigma_{xy}\approx9 mm$ \cite{nygren2009high}. We use 1.0 mm, 9.0 mm standard deviation for transverse diffusion and 1.0 mm, 2.0 mm, 5.0 mm, 9.0 mm standard deviation for longitudinal diffusion as sampling points to analyze the effect of diffusion on the proposed model.

\section{Deep learning method}
\label{section:Deep learning method}

Artificial neural networks have been introduced into the field of experimental physics (especially high-energy physics) since 1988 \cite{denby1988neural,peterson1989track}, for both online triggering and offline data analysis. However, a recent development called "deep learning" has just been acknowledged by the community of physics, and its application is restricted to a few specific domains \cite{baldi2014searching,de2016jet,racah2016revealing}. In the experiment of searching for neutrinoless double-beta decay, we can collect data in the detector and reconstruct the 3-dimensional topological signature of candidate events. An important issue is how to select signal events from these candidates while significantly rejecting background events. In \cite{Renner:2016trj}, a 2D convolutional neural network was used for background rejection in the NEXT experiment. The input of the network accepted three coordinate plane projections of the 3D signature, which acted in the same way as the three color channels in an RGB image. This work achieved the competitive result in this field; nevertheless, some issues are still left open: how to customize the Convolutional Neural Network (CNN) to fit into the setting of the experiment, and how to make use of the 3D signature more thoroughly. Here, some basic theories of our approach are explained in Section \ref{subsection:Deep learning and CNN}, and more advanced structures are analyzed in detail in Section \ref{subsection:3D CNN} and Section \ref{subsection:3D ResNet}.

\subsection{Deep learning and CNN}
\label{subsection:Deep learning and CNN}

Early neural networks contained only one hidden layer, and their practicality was supported by the universal approximation theorem \cite{hornik1989multilayer,cybenko1989approximation}. In recent years, with the advancement in hardware and software, deep neural networks have gained extended popularity in image classification \cite{krizhevsky2012imagenet}, speech recognition \cite{hinton2012deep}, natural language processing \cite{shen2014learning} and many other scientific fields \cite{kayala2011learning,magnan2014sspro}. Meanwhile, mathematical inferences demonstrated that increasing depth could not only gain power in feature extraction, but also reduce the number of parameters needed to approximate some specific functions \cite{pascanu2013number,montufar2014number}.

Two supporting theories behind this trend are deep feedforward networks (also called multi-layer perceptrons) and the back propagation algorithm. The simplest unit of a neural network is called a \emph{neuron}. It computes the dot product between an input vector and a trainable weight vector, adds a trainable bias, and achieves nonlinearity through an activation function. Many similar neurons act on the same input vector and use different weights and biases to construct a "layer". Furthermore, if we take the outputs of the previous layer as inputs and add a new layer of neurons, we can get one more layer, and so on. Finally, the essential part of a deep feedforward network is built in this way. The intermediate layers in such a network are called hidden layers. Usually we need several hidden layers to form a deep model, but sometimes we need even more \cite{szegedy2015going,he2016deep}.

To train such a feedforward network, the standard method is the back propagation algorithm. It uses the loss function which indicates the discrepancy between the network output and the expected output (usually from human labeling or simulation) to update parameters in each of the hidden layers in a reverse manner. The loss function can take various forms; however, the most commonly used forms are already conventions. For a classification task, usually we use the cross entropy along with the softmax function \cite{krizhevsky2012imagenet}; for a regression task, we use the mean square error and its derivatives \cite{girshick2015fast}.

In the past few years, Convolutional Neural Networks (CNN) have been designed to solve image recognition problems and achieved significantly better results \cite{krizhevsky2012imagenet,szegedy2015going,he2016deep} than traditional methods. The \emph{convolution} in machine learning libraries is an operation which calculates the discrete cross-correlation between two n-dimensional arrays in an element-wise manner. Based on the translation-invariance of the image, parameters in hidden layers can be greatly reduced by sharing parameters between different locations in the image. In practice, we use a set of m $\times$ n weights and a bias to form a \emph{filter} or \emph{kernel}. Typically the image has more than one channel, so we need m $\times$ n weights for each channel. By applying these weights to the input and adding the bias, we can get one output per location per filter. Moving the filter in horizontal and vertical directions in a minimum unit called stride, we can get a new 2D plane. Often we need a plenty of filters, each of which performs independently. So the final output (i.e., a feature map) is quite the same as the input: 2D planes with many channels, each of which corresponds to one filter. This operation can be executed iteratively until we get enough hidden layers for our model. To reduce the size of the feature map, we can use pooling layers. The most commonly used pooling method is the max pooling: computing the max value in an m $\times$ n region of the feature map, moving the region by a stride and across all channels, and finally getting a reduced feature map with the compressed height and width but the same channels.

\subsection{3D CNN}
\label{subsection:3D CNN}

\begin{figure}[htbp]
	\subfigure[3D convolution layer]{			
		\includegraphics[width=0.48\textwidth]{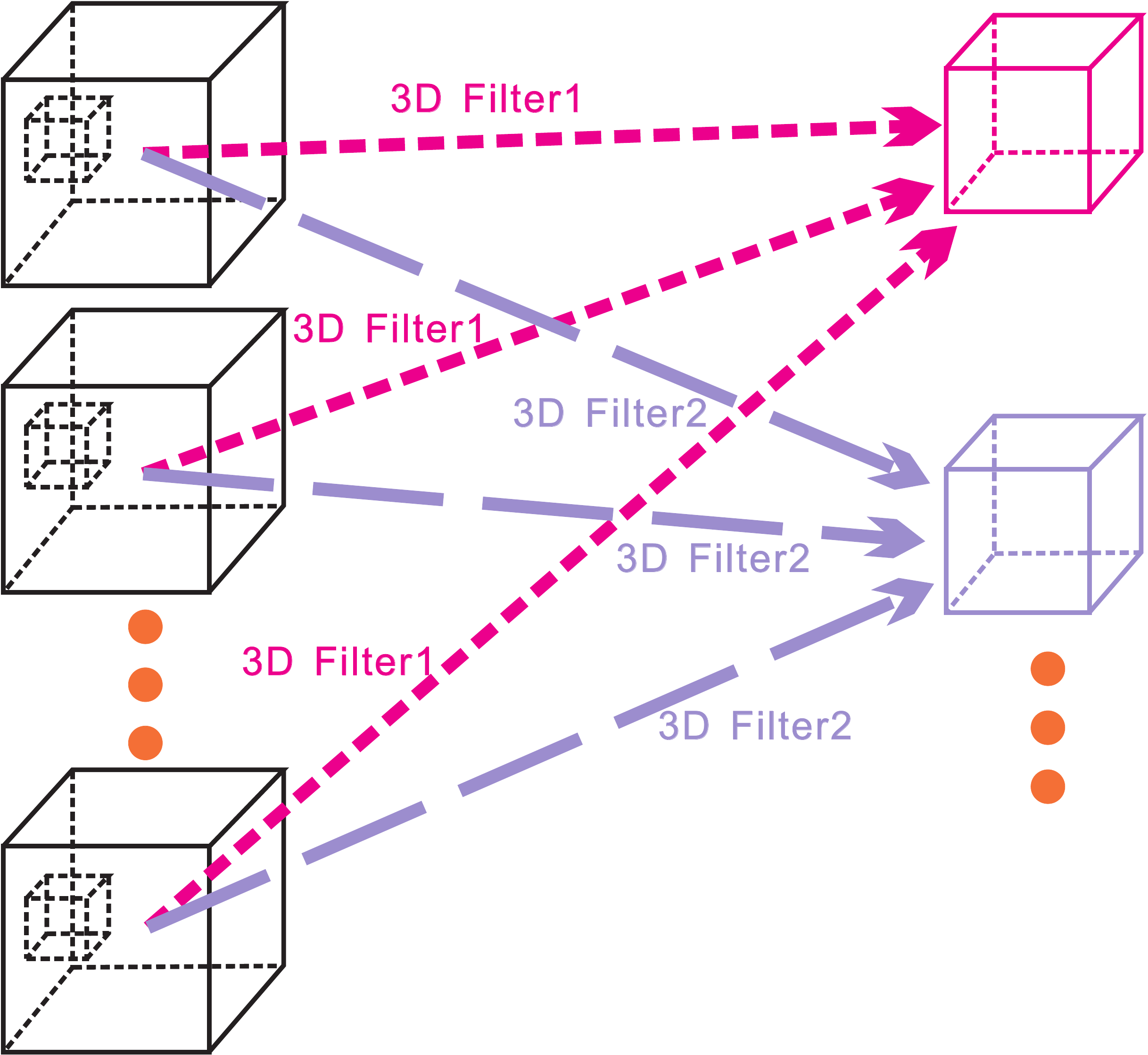}}		
	\subfigure[3D max pooling layer]{
		\includegraphics[width=0.48\textwidth]{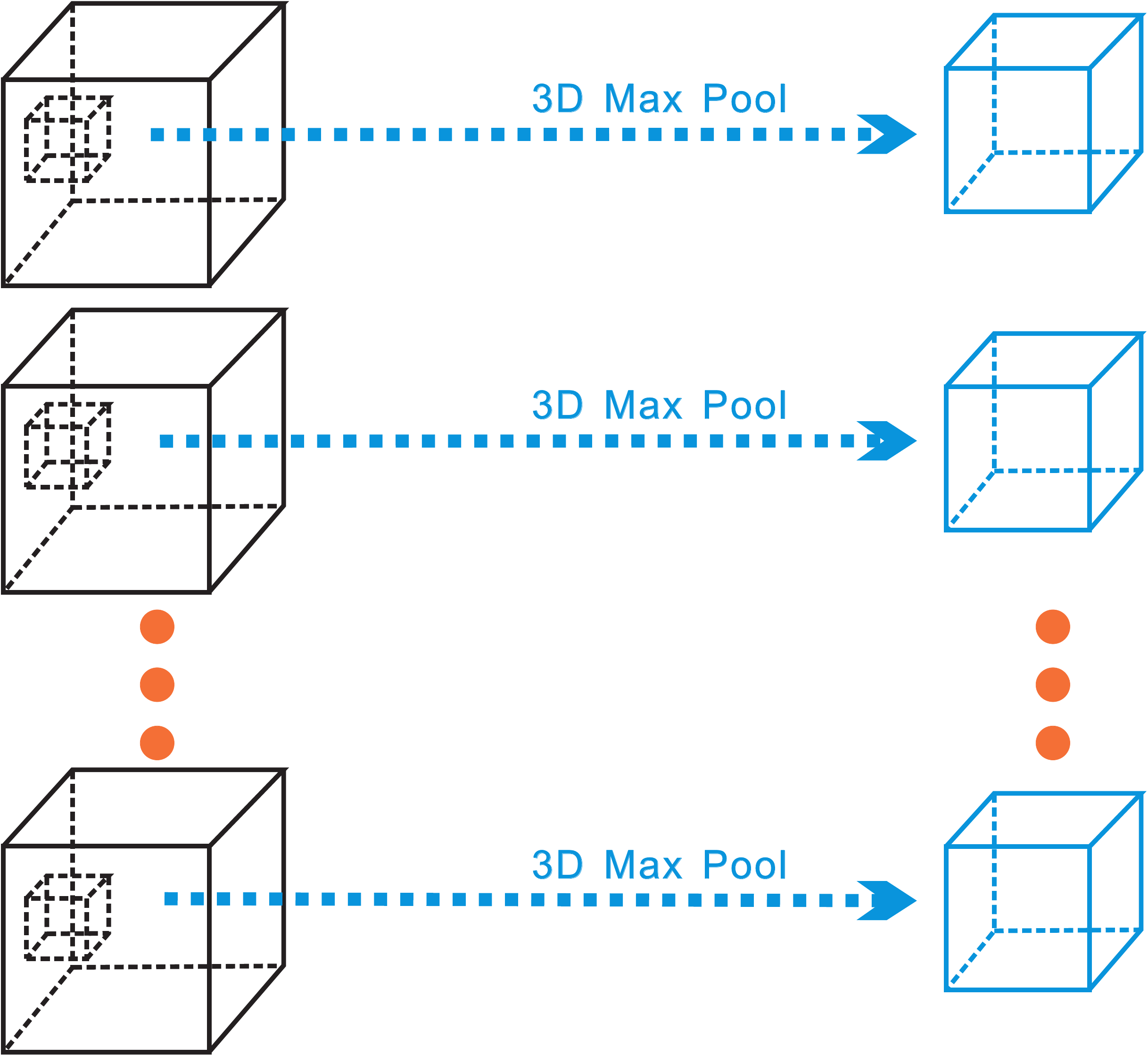}}
	\caption{\label{fig:3D CNN} A schematic diagram of two common layers in a 3D convolutional neural network.}
\end{figure}

In the setting of our experiment, original reconstructed events are 3D spatial distributions of charge density. While using coordinate plane projections is a good compromise between the data format and deep learning conventions, we still want to fully exploit the underlying information carried by the voxels ("voxel" is a terminology for the element in a 3D array). There are several advantages in this method. First, 3D relative positions can be preserved and utilized by our deep learning model. Second, 3D analysis makes it possible to use more effective data preprocessing methods to facilitate feature extraction, which will be discussed in Section \ref{subsection:Data preprocessing}. Last, 3D voxels have specific physical implications, such as total energy, which could be spoiled by the projection operation.

Prior to our work, 3D Convolutional Neural Networks (3D CNN) have been designed for object classification \cite{maturana2015voxnet}, upright orientation regression \cite{liu2016upright}, medical image analysis \cite{kleesiek2016deep}, etc. Analogous to the 2D CNN, the basic theory for the 3D CNN is direct and intuitive. However, because datasets to which 3D CNNs could be applied are limited in general, 3D CNNs are not widely used and their application scenarios are still under exploration. In our $0\nu\beta\beta$ experiment, a considerable amount of 3D pictures will be produced by the triggering and reconstruction process. With the data available for machine learning, how to build an effective discriminative model selecting signal events in a background dominated circumstance becomes a pressing demand. This motivates us to put forward new CNN structures.

The major difference between 2D and 3D CNNs lies in the definition of the convolution operation. A schematic diagram is shown in Fig. \ref{fig:3D CNN}. Typically, the input layer (which may be a feature map) is a 4-dimensional tensor (depth, height, width, channel), and depth is an added dimension. Hence the size of weights for each channel in each filter should be m $\times$ n $\times$ k. When we move the filter across the input layer, we can move in three directions, along three axes. A filter also contains a set of weights and a bias, and applying them to the input layer will result in a new 3D cuboid. A plenty of filters will finally produce a new 4D tensor, which is ready for the next 3D convolution. For 3D pooling, the principle is the same as in the 2D case. We reduce the size in the first three dimensions and leave the channel unchanged.

Adding a new dimension can be implemented systematically, but it has brought about some questions. Because the depth is a new multiplier, the actual size of each dimension can not be too big ($\sim$65). Meanwhile, a new dimension also means a new degree of freedom, which needs more convolution operations. So a practical model has limits on the input 3D picture and total parameterized layers.

\subsection{3D ResNet}
\label{subsection:3D ResNet}

When exploring network structures to get better classification results, the advantage of the depth of the whole network is shown. In order to train a deeper neural network, some normalization techniques \cite{krizhevsky2012imagenet,ioffe2015batch} are necessary to avoid the problem of vanishing/exploding gradients. However, even though deep neural networks can converge by using normalization, a degrading problem can still prevent the network from achieving better results, whether in the training process or in the evaluation.

In \cite{he2016deep}, a simple and effective improvement to the network structure has been proposed, which can support much deeper networks without causing degradation. The delicate building block is implemented by a main path consisting of several parameterized layers, and a shortcut connection directly linking the input and the output of the main path. In principle, the shortcut connection only performs the identity mapping, i.e. passing the input to the output without changing anything. As a result, the main path only needs to learn the incremental value on the basis of the identity mapping, and this can greatly alleviate the burden of stacked layers. This method has been proven to be very effective in building very deep neural networks. The neural network using this structure is called ResNet.

As ResNet was so successful in 2D CNNs for image classification and detection tasks, it may be very natural to generalize it to the 3D CNN case. We call it "3D ResNet"; the convolution operates on 3D voxels and we use ResNet building blocks to replace some of the 3D convolution layers. The ResNet blocks can fit into the original model smoothly, and we can use the same optimizing algorithm to train the network.

We have made two alterations to the original ResNet block: for one thing, all 2D convolution layers in the main path are replaced by 3D convolution layers, which is internally supported by our software framework; for another, we expand the 2D batch normalization to the 3D case, where depth, height and width along with minibatch are involved in a Gaussian normalization. We use an architecture similar to the "bottleneck" architecture \cite{he2016deep}, with some modifications in order to compare to other 3D CNNs. In this architecture, there are three parameterized layers in the main path, which has an advantage over two parameterized layers.

\section{Network architectures}
\label{section:Network architectures}

\begin{table}[htbp]
	\centering
	\caption{\label{tab:Network architectures} Network architectures.}
	\begin{threeparttable}
		\small
		\begin{tabular}{l|l|llll}
			\hline & 2D\_toy & 2D\_base & 3D\_layer5 & 3D\_layer7 & 3D\_Res \\ 
			\hline (input) & 80$\times$80 & 299$\times$299$\times$3 & 20$\times$20$\times$20 & 20$\times$20$\times$20 & 20$\times$20$\times$20 \\
			1 & Conv(5,2,32)\tnote{a} & Conv(11,4,64) & Conv(3,1,32) & Conv(3,1,16) & Conv(3,1,16) \\ 
			2 & Conv(5,2,32) & --- & --- & Conv(3,1,16) & Res(3,1,16)\tnote{f} \\ 
			  & Pool(3,2)\tnote{b} & Pool(3,2) & Pool(3,2) & Pool(3,2) & Pool(3,2) \\ 
			3 & Conv(3,1,64) & Conv(5,1,64) & Conv(3,1,64) & Conv(3,1,32) & Res(3,1,32) \\ 
			4 & Conv(3,1,64) & --- & --- & Conv(3,1,32) & Res(3,1,32) \\ 
			  & Pool(3,2) & Pool(3,2) & Pool(3,2) & Pool(3,2) & Pool(3,2) \\ 
			5 & Conv(3,1,128) & Conv(3,1,64) & Conv(3,1,64) & Conv(3,1,64) & Res(3,1,64) \\ 
			  & Pool(3,2) & Pool(3,2) & Pool(3,2) & Pool(3,2) & Pool(3,2) \\ 
			6 & FC(512)\tnote{c} & FC(1024) & FC(512) & FC(512) & FC(512) \\ 
			  & Dropout(0.5)\tnote{d} & Dropout(0.5) & Dropout(0.5) & Dropout(0.5) & Dropout(0.5) \\ 
			7 & FC(2) & FC(2) & FC(2) & FC(2) & FC(2) \\ 
			  & Softmax(2)\tnote{e} & Softmax(2) & Softmax(2) & Softmax(2) & Softmax(2) \\ 
			(for training) & Cross entropy & Cross entropy & Cross entropy & Cross entropy & Cross entropy \\ 
			\hline 
		\end{tabular}
		
		\begin{tablenotes}
			\footnotesize
			\item[a] Conv(x, y, z) means a convolution layer with filter size x, stride y and output channels z.
			\item[b] Pool(x, y) means a max pooling layer with filter size x and stride y.
			\item[c] FC(x) means a fully-connected layer with length x.
			\item[d] Dropout(x) means a Dropout layer with dropout ratio x when training.
			\item[e] Softmax(x) means a softmax layer with x output probabilities.
			\item[f] Res(x, y, z) means a ResNet building block with filter size x, stride y and output channels z.				
		\end{tablenotes}
	\end{threeparttable}
\end{table}

Table \ref{tab:Network architectures} is an overview of our network architectures. Generally, we use 5 different models: 2D\_toy, 2D\_base, 3D\_layer5, 3D\_layer7 and 3D\_Res. The 2D\_toy is used to work on the synthetic data of the toy model in Section \ref{section:A toy model demonstration}. The 2D\_base is mainly used for the granularity simulation (Section \ref{subsection:Granularity simulation}) and in comparison with 3D CNNs for the 3D network simulation (Section \ref{subsection:3D network simulation}). All 3D CNNs are involved in the 3D network simulation, and 3D\_Res is chosen to work on the diffusion and noise simulation (Section \ref{subsection:Diffusion and noise simulation}) in more realistic conditions. 

We denote layers in a concise manner, and the interpretation of our symbolic language can be seen in the footnotes of Table \ref{tab:Network architectures}. For simplicity, Rectified Linear Units (ReLU) \cite{nair2010rectified} after each parameterized layer are omitted, but they are essential parts of the networks for proper operation. Because the inputs to the networks are square or cubic in shape, we choose the same size for depth, height or width, and this can be denoted in only one number. Notice that we use the same symbol for both the 2D CNN and the 3D CNN, but the actual filter in the operation of convolution or max pooling is different: the former is two-dimensional and the latter is three-dimensional. For example, if the filter size is 3, the 2D CNN has filters of 3 $\times$ 3 and the 3D CNN has filters of 3 $\times$ 3 $\times$ 3.

\begin{figure}[htbp]
	\centering
	\includegraphics[width=0.9\textwidth]{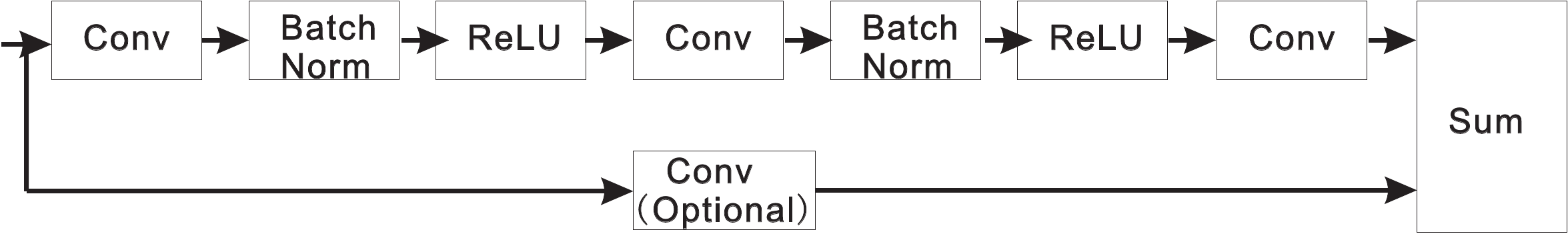}
	\caption{\label{fig:3D ResNet building block} The 3D ResNet building block.}
\end{figure}

In 3D\_Res, we use ResNet building blocks to replace all the convolution layers except the first one in the 3D\_layer7. This has a significant gain in the total depth of the network. More precisely, the ResNet building block is not a "layer", and its inner structure can be seen in Fig. \ref{fig:3D ResNet building block}. In this figure, we can see that there are three convolution layers in the main path, followed by batch normalization layers and ReLUs. The filter size of successive Conv layers, the stride of the first Conv layer and the output channels of the last Conv layer are given in the table. Other parameters are fixed or dependent on previous layers. The optional convolution layer is used to match the output channels in order to sum up the main path and the shortcut connection in an element-wise manner.

\section{A toy model demonstration}
\label{section:A toy model demonstration}

Before going into the details of the simulation in realistic conditions, some analysis is beneficial to understand the discrimination power of the neural network model. In \cite{Renner:2016trj}, the authors ran toy Monte Carlo (MC) simulations to demonstrate the reason why a significant amount of events were misclassified by the CNN. Although comparisons between the toy MC and realistic data were made in the paper, the ablation simulation could only demonstrate how physical interactions degrade the performance of the CNN.

In this part, we want to address more abstract questions regarding the origin of the discrimination power. While signal events are different from background events in their signature patterns, we cannot achieve 100\% accuracy, because some samples are intrinsically inseparable due to the probabilistic model. If samples are generated according to the probabilistic model, we have "corner cases" of signal events that look like background events, and vice versa. This confusing nature of the probabilistic model introduces the Bayes error, the innate error rate when the probabilistic model is well defined. In other words, there exists a theoretical upper limit that cannot be crossed by the neural network model.

However, using good machine learning techniques, such as the CNN, we can hope that our model could achieve an accuracy as close to the theoretical limit as possible. Since the input of the CNN is not the probabilistic model itself, but its projection onto a high-dimensional space, there is a loss of information. We want the CNN to utilize the remaining information thoroughly so that no significant drop of performance is observed when compared to the theoretical limit.

In the particular problem of neutrinoless double beta decay, the unique signature of signal events has the following features:

\begin{enumerate}
	\item It starts with a single point where the double beta decay happens. Two electrons are shot to two distinct directions, leaving an initial angle.
	\item The tracks of the two electrons suffer from multiple scattering, which makes them continuous but curved.
	\item At the two ends of the tracks, a considerable amount of energy is deposited (Bragg peak), and the ionization becomes significant.
\end{enumerate}

In contrast, the signature of background events is different in the following ways:

\begin{enumerate}
	\item Since only one electron is involved in the physical process, no initial angle exists in the track.
	\item The single electron has higher energy, so the curvature of its track is different from double beta events.
	\item There is a great density of energy depositions at only one end of the track.
\end{enumerate}

\begin{figure}[htbp]
	\centering
	\subfigure[Signal]{			
		\includegraphics[width=0.40\textwidth]{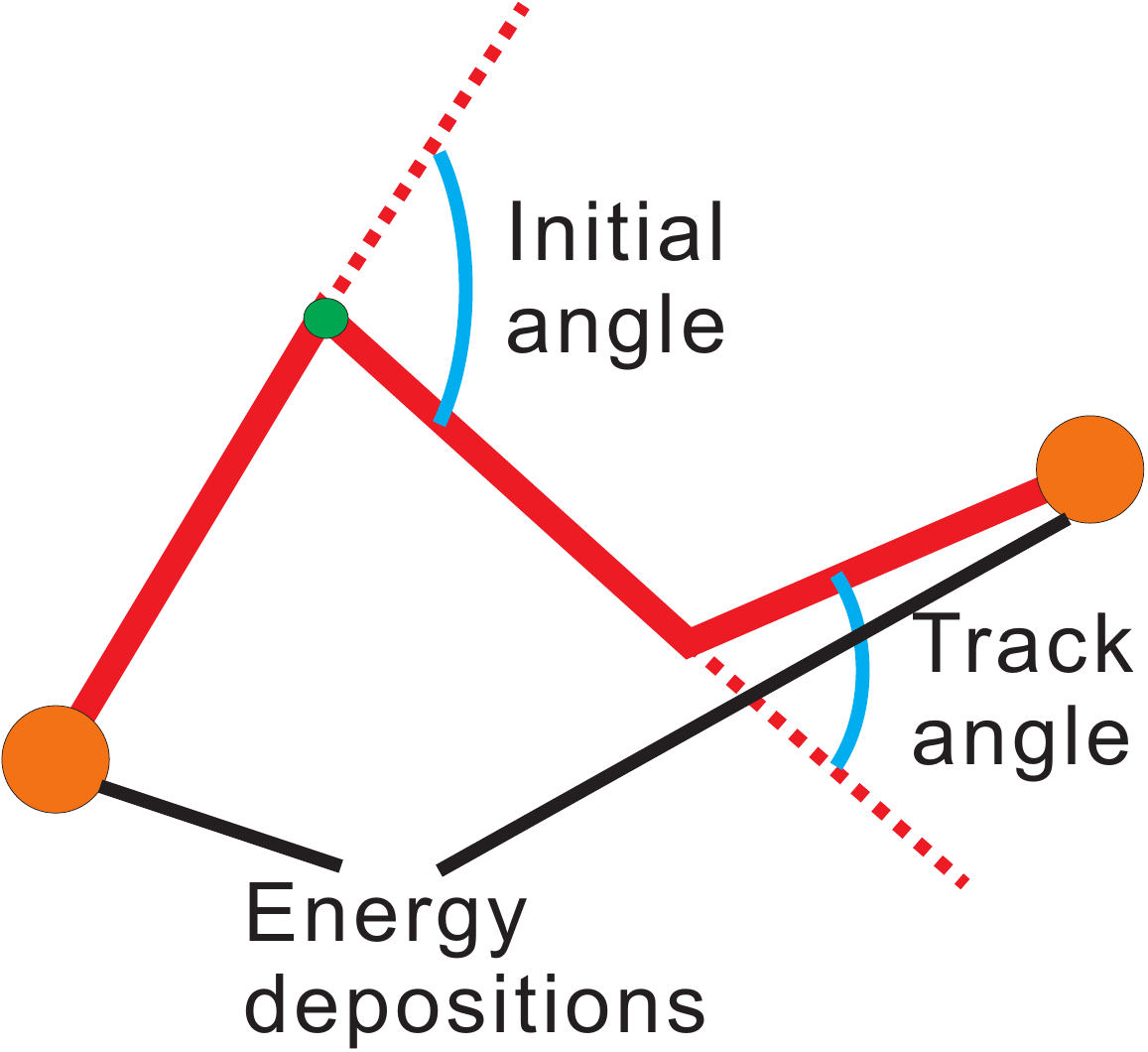}}		
	\subfigure[Background]{
		\includegraphics[width=0.45\textwidth]{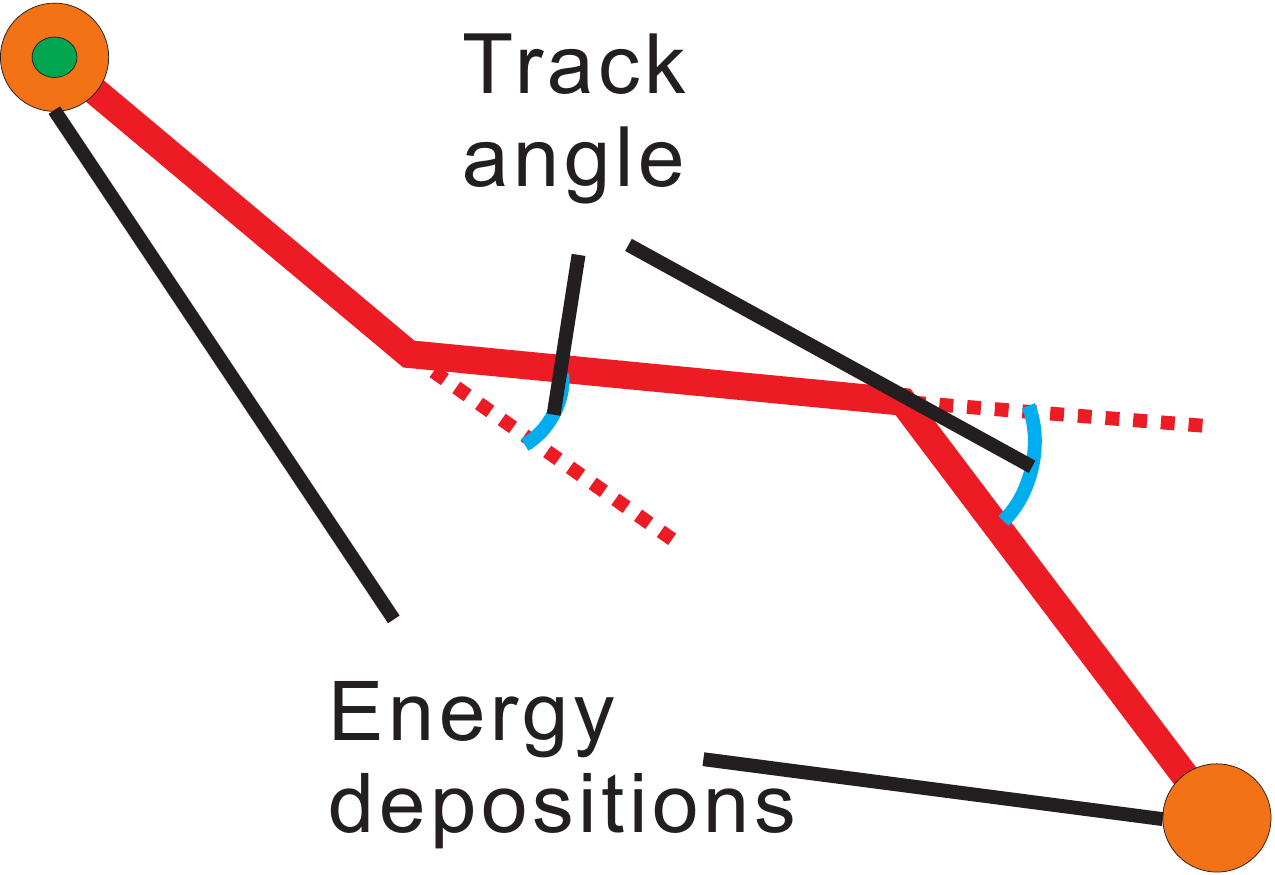}}
	\caption{\label{fig:Toy model} The three-segmented toy model for signal and background events.}
\end{figure}

To deal with the problem, we introduce our \emph{segmented chain with energy depositions} toy model (Fig. \ref{fig:Toy model}). A detailed description of our toy model is stated in Appendix \ref{section:Details of the toy model}. In short, first we build the probabilistic model of signal events and background events with the same number of segments and two depositions. Second, we calculate the theoretical upper limit (one minus the Bayes error) using the probabilistic model. Third, we generate 2D images of signal events and background events using the same parameters as those used in the calculation. Last, we train the CNN model, test the CNN model on 2D images and compare the results to the theoretical limits.

\begin{table}[htbp]
	\centering
	\caption{Results of the toy model simulation and theoretical limits.}
	\label{tab:Toy model and limits}
	\begin{threeparttable}
		\begin{tabular}{lll}
			\hline Feature of the toy model & Accuracy of the CNN (2D\_toy) (\%) & Theoretical limit (\%) \\ 
			\hline 2 segments, no depositions & 70.5 & 71.2  \\
			       3 segments, no depositions & 79.1 & 80.8  \\
			       only depositions & 88.1 & 91.1  \\
			       2 segments, with depositions & 88.2 & 93.1 \\  
			       3 segments, with depositions & 90.6 & 95.7 \\
			\hline mixing model\tnote{a} & 89.3 & 94.4 \\
			\hline 
		\end{tabular}
	
		\begin{tablenotes}
			\footnotesize
			\item[a] the mixing model represents the mixture of 2 segments and 3 segments with depositions. It is closer to the actual physical process than other conditions.
		\end{tablenotes}

	\end{threeparttable}
\end{table}

The simulation result is shown in Table \ref{tab:Toy model and limits}. In the table, a total of 6 conditions are analyzed. Some of the conditions (the first three rows) don't represent the actual physical process; we analyze them only for decomposing different factors. In general, the performance of the CNN is consistent with the theoretical limits. In the first two rows, the results achieved by the CNN are very close to the theoretical limits (0.7\% and 1.7\% gap), which demonstrates that CNNs can utilize the angle information in the conjunction of the segments very well. In the third row, the result of "only depositions" has a gap of 3\%, which is slightly larger, but also shows promising performance. In the fourth row, the fifth row and the sixth row, the results of complete toy models have relatively large gaps (4.9\%, 5.1\% and 5.1\% gap). If we compare the fourth row and the third row, we can see almost no improvements in the accuracy of the CNN. This is probably because in the probabilistic model, we can always combine the information perfectly, which is not the case for CNNs. This also shows a direction for future improvements of the discrimination by machine learning methods.

\section{Data simulation and preprocessing}
\label{section:Data simulation and preprocessing}

\subsection{Data simulation}
\label{subsection:Data simulation}

Based on the instrument specifications, the simulation process generates a considerable amount of signal and background events using the Monte Carlo method. For $0\nu\beta\beta$ events, the initial kinematics of electrons in double beta decays is calculated by the DECAY0 package (see their latest paper \cite{ponkratenko2000event}). According to the momentum distribution defined by the DECAY0, we randomly generate signal events with their initial states, then feed the initial states to the Geant4 \cite{agostinelli2003geant4} simulation toolkit for simulating the passage of electrons in the \ce{^{136}Xe} high pressure gas. For background events, electrons with similar energy (2.458 MeV) are created and cast into the same geometry as signal events in Geant4.

We employ a square pattern instead of the hexagonal pattern for simplicity. Each event is a tensor $(ed, xp, yp, zp)$, in which $ed$ stands for the amount of energy deposited, and $(xp, yp, zp)$ stands for the 3D coordinates of the energy depositions. $0\nu\beta\beta$ events are randomly created in the active region of the detector, while $beta$ events with similar total energy are created in the same manner. In the next step, the energy depositions are converted to ionization charges using a W-value of 24.8 eV \cite{platzman1961total,alvarez2013near} and considering the Fano factor in \ce{Xe} gas ($0.13-0.17$) \cite{bolotnikov1997spectroscopic,alvarez2013near}.

After that, the diffusion of charges is considered, and the noise is added to the value. The mechanism we use to generate the diffusion is described below. First, we use the transverse diffusion factor and the longitudinal diffusion factor as singular values to produce a covariance matrix. Second, we calculate the Cholesky decomposition of the covariance matrix. Third, we use random origins and the output matrix of the second step to produce a three dimensional Gaussian distribution. Last, we add the Gaussian distribution to the 3D coordinates of each ionization charge to account for the diffusion effects. For noise, we use the Gaussian distribution with a fixed deviation to generate noise values centering at zero, and add them to voxels (See Section \ref{subsection:Diffusion and noise simulation} for more explanation).

Finally, the charges are printed out in the order of three dimensions. These 3D discrete points can be fed to later preprocessing and used as the input of neural network models (Section \ref{section:Network architectures}).

\subsection{Data preprocessing}
\label{subsection:Data preprocessing}

\paragraph{2D projection}

\begin{figure}[htbp]
	\centering
	\subfigure{			
		\includegraphics[width=0.95\textwidth]{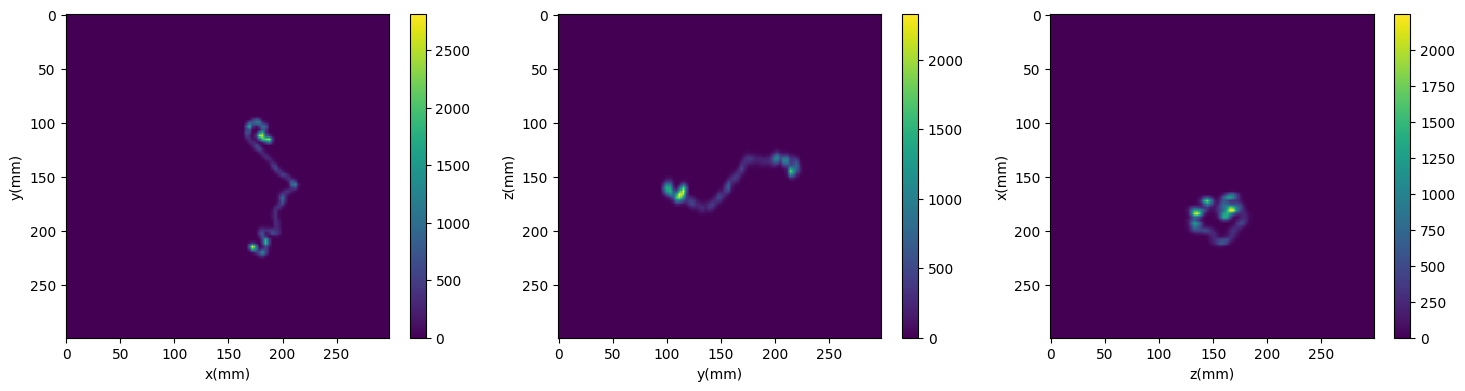}}		
	\subfigure{
		\includegraphics[width=0.95\textwidth]{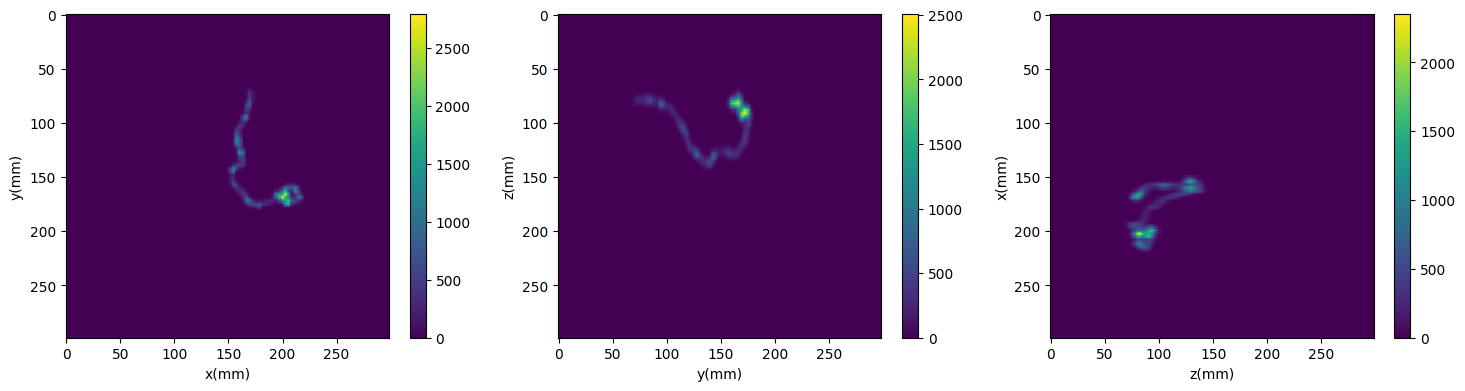}}
	\caption{\label{fig:2D projection} Visualization of 2D coordinate plane projections after the boundary test and resize. (Top) Three channels of a $0\nu\beta\beta$ event sample. (Bottom) Three channels of a $beta$ event sample.}
\end{figure}

Inspired by \cite{Renner:2016trj}, we generate coordinate plane projections (Fig. \ref{fig:2D projection}) from 3D signatures as our dataset for the granularity simulation. The original picture is a 250 $\times$ 250 16-bit png file, with three color channels corresponding to xy, yz and zx projections, and the granularity is 2 mm $\times$ 2 mm $\times$ 2 mm. Through visualization, we find that in most pictures the active (nonzero) pixels are restricted in the center region, which is small compared to the actual size. For making the 2D CNN work with its maximum ability, we run a boundary test to select those pictures with active pixels within the center 100 $\times$ 100 region (which corresponds to the size of 200 mm $\times$ 200 mm $\times$ 200 mm) and resize it to 300 $\times$ 300, which is the designated input size of our 2D CNN model. To perform the granularity simulation, we need to simulate pictures with different granularities. The strategy is to use the area average method to shrink the size of the picture and then use the nearest neighbor interpolation to amplify the picture. It works well, in the ideal condition (no noise); the 3D charge distribution is blurred in a mosaic way without changing the total charge in any of the coordinate plane projections.

\paragraph{3D cuboid}

\begin{figure}[htbp]
	\centering
	\subfigure{			
		\includegraphics[width=0.23\textwidth]{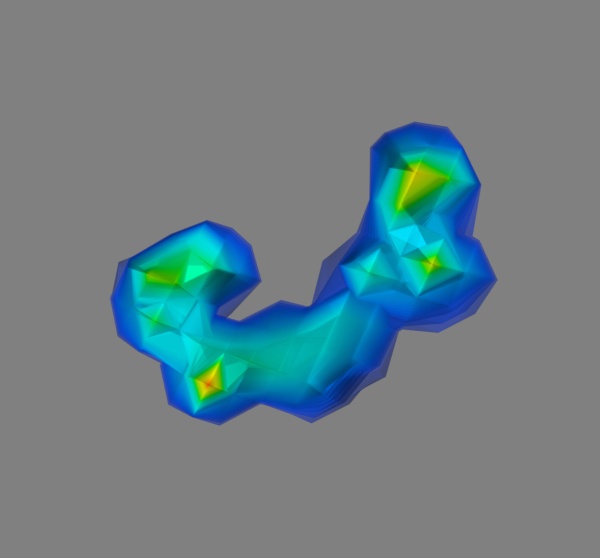}}		
	\subfigure{
		\includegraphics[width=0.23\textwidth]{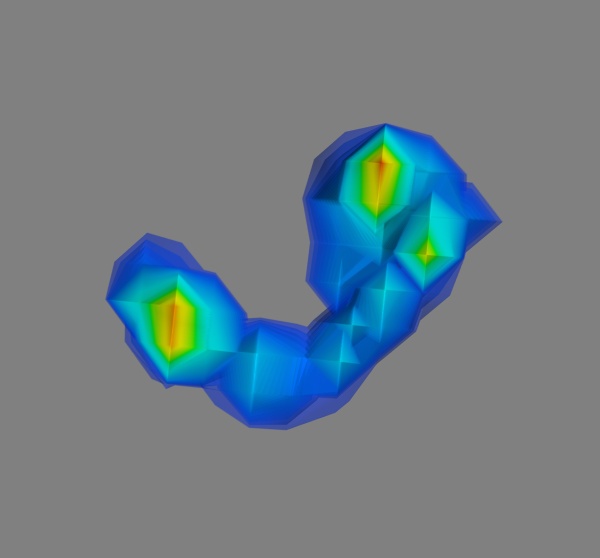}}
	\subfigure{			
		\includegraphics[width=0.23\textwidth]{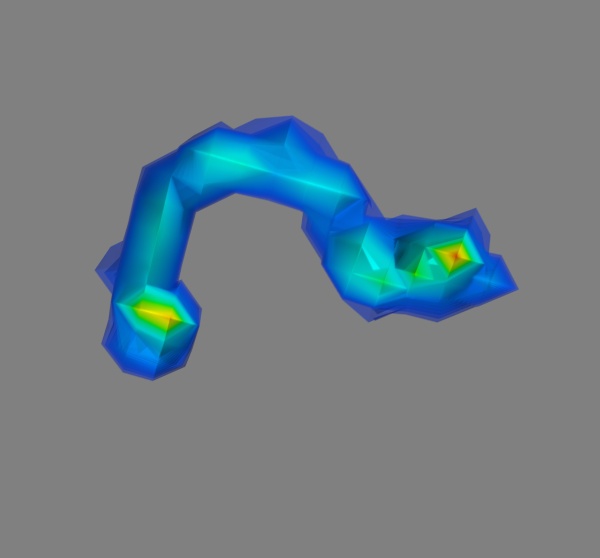}}		
	\subfigure{
		\includegraphics[width=0.23\textwidth]{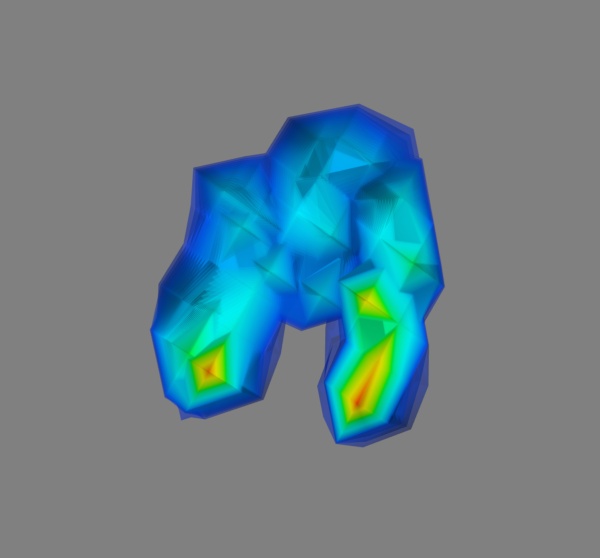}}
	\subfigure{			
		\includegraphics[width=0.23\textwidth]{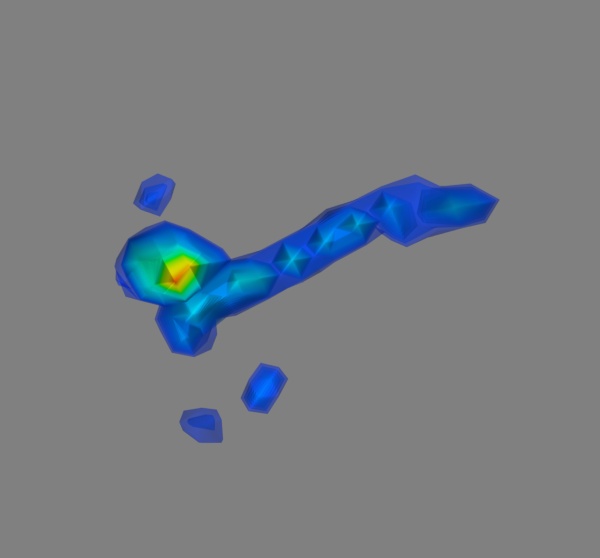}}		
	\subfigure{
		\includegraphics[width=0.23\textwidth]{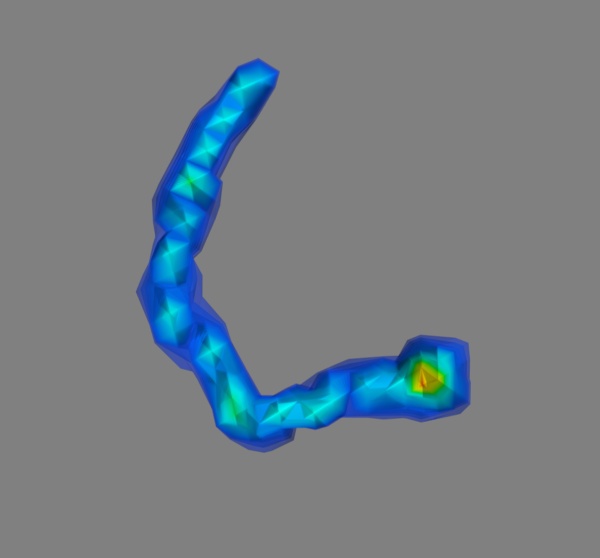}}
	\subfigure{			
		\includegraphics[width=0.23\textwidth]{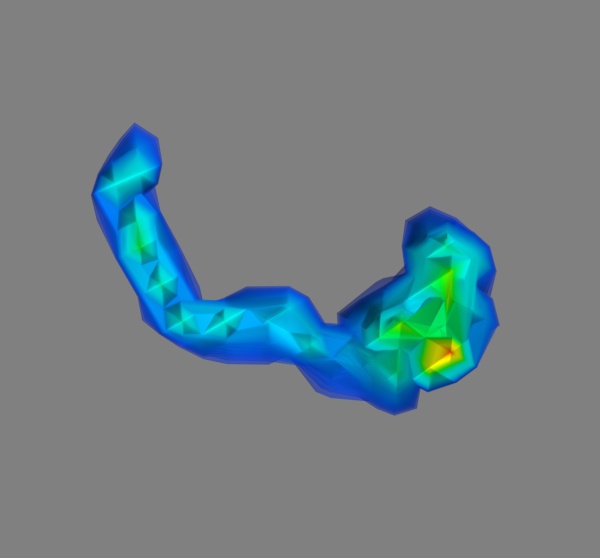}}		
	\subfigure{
		\includegraphics[width=0.23\textwidth]{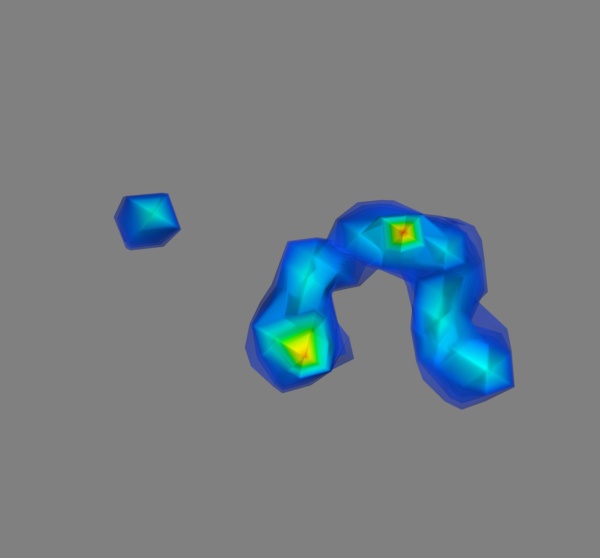}}
	\caption{\label{fig:3D cuboid} Visualization of 3D cuboids after preprocessing. (Top) Four samples of $0\nu\beta\beta$ events. (Bottom) Four samples of $beta$ events.}
\end{figure}

For the purpose of studying the problem in depth, we generate 3D cuboid pictures (Fig. \ref{fig:3D cuboid}) and use them in combination with the 3D CNN and the 3D ResNet. The size of the 3D cuboid is 50 $\times$ 50 $\times$ 50 and the granularity is fixed to 8 mm $\times$ 8 mm $\times$ 8 mm. Visualizing the 3D picture may not be as clear as the 2D picture, but statistical work shows that only about 100 voxels in a 3D picture are active, which may be too scarce ($\sim$1/1250) for 3D CNNs to learn directly. Since the spatial information is preserved, we can use an algorithm to select a 20 $\times$ 20 $\times$ 20 region fully surrounding the active region and locate it in the center. This reduces the size of the cuboid to 160 mm $\times$ 160 mm $\times$ 160 mm and increases the ratio of active voxels to about 1/80. When noise is presented, this strategy is still valid, for voxels apparently above noise are being considered and put in the center. This strategy establishes a threshold and wipes out voxels which have an amount of charge below the threshold. Since the region of active voxels is relatively limited, to exclude some voxels with non-zero charge has a minor effect on the integrity of the input cuboid.

\paragraph{Simulation data loss}

In both 2D and 3D data preprocessing, we select samples subject to some constraints, and this results in a loss of data. For the 2D projection, loss rates are 13\% and 44\% for $0\nu\beta\beta$ events and $beta$ events, respectively. For the 3D cuboid, the loss rate is about 10\% for both $0\nu\beta\beta$ events and $beta$ events. Because the failed samples mostly contain "energy isolated islands" caused by bremsstrahlung, this loss is acceptable in the current stage of research for concentrating on the pattern difference between signal events and background events.

\section{Simulation studies}
\label{section:Simulation studies}

In the following parts, we run simulations from three different perspectives: (1) how the granularity affects the performance of the neural network model; (2) how different architectures of the neural network model differ in their discrimination power; and (3) how diffusion and noise degrades or improves the performance of our model.

In the simulation studies, we experienced substantial fluctuations in the performance metrics of CNNs (especially the accuracy metric). To reduce the randomness of the result, if the iteration steps are not specified, we plotted the graph or calculated the mean value and the standard deviation in the converged region on a separate validation dataset. Otherwise, we used the validation set as the test dataset to give one-shot results at specific iteration steps.

\subsection{Granularity simulation}
\label{subsection:Granularity simulation}

\begin{figure}[htbp]
	\centering
	\includegraphics[width=0.9\textwidth]{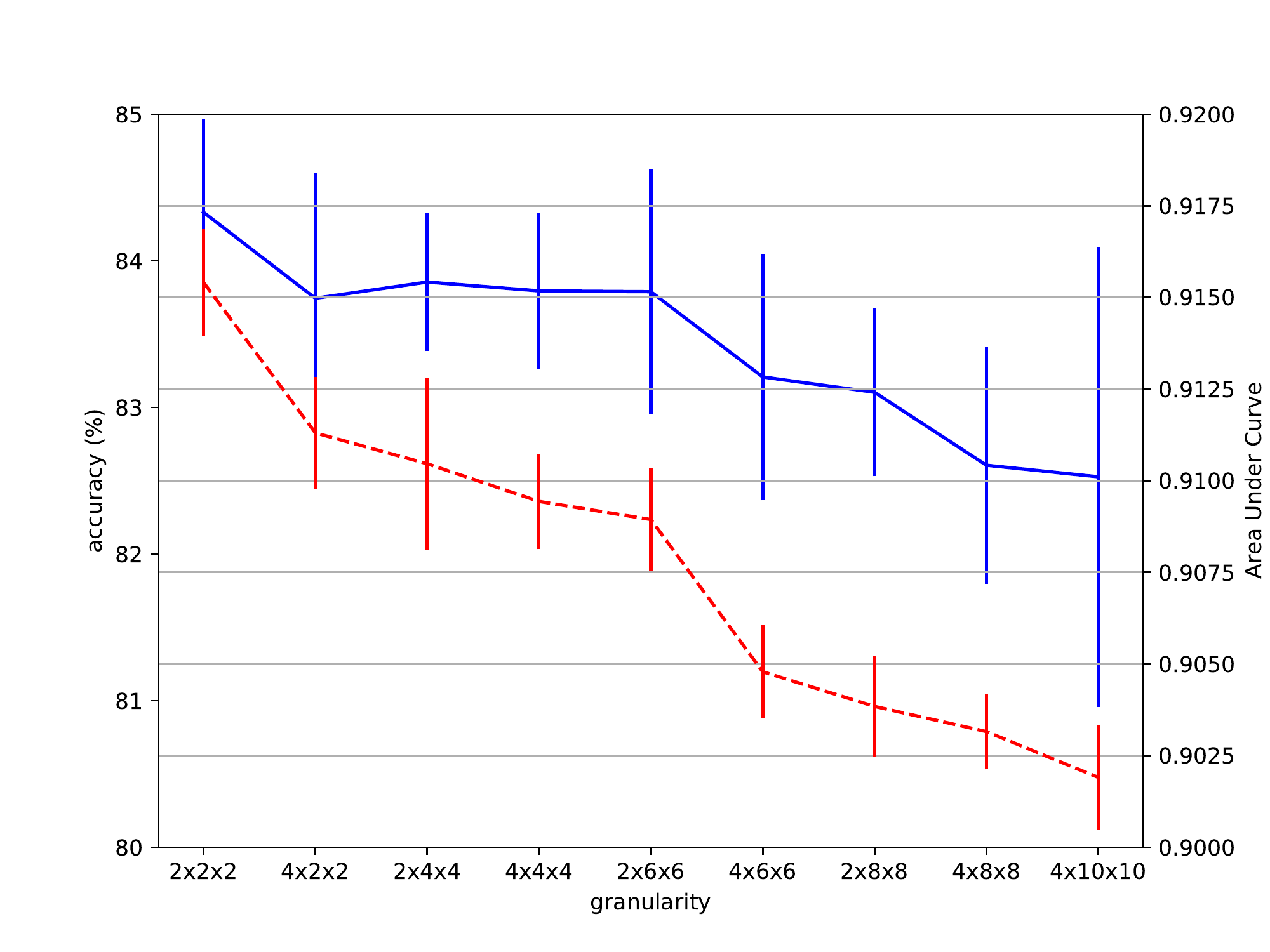}
	\caption{\label{fig:Accuracy and AUC influenced by granularities} Accuracy and AUC (Area Under Curve) influenced by different granularities. The label of the horizontal axis represents the values in three dimensions (x, y, z in order). The mean value and the standard deviation are calculated in the converged region. Solid line: accuracy in percentage. Dashed line: Area Under Curve.}
\end{figure}

\begin{figure}[htbp]
	\centering
	\includegraphics[width=0.9\textwidth]{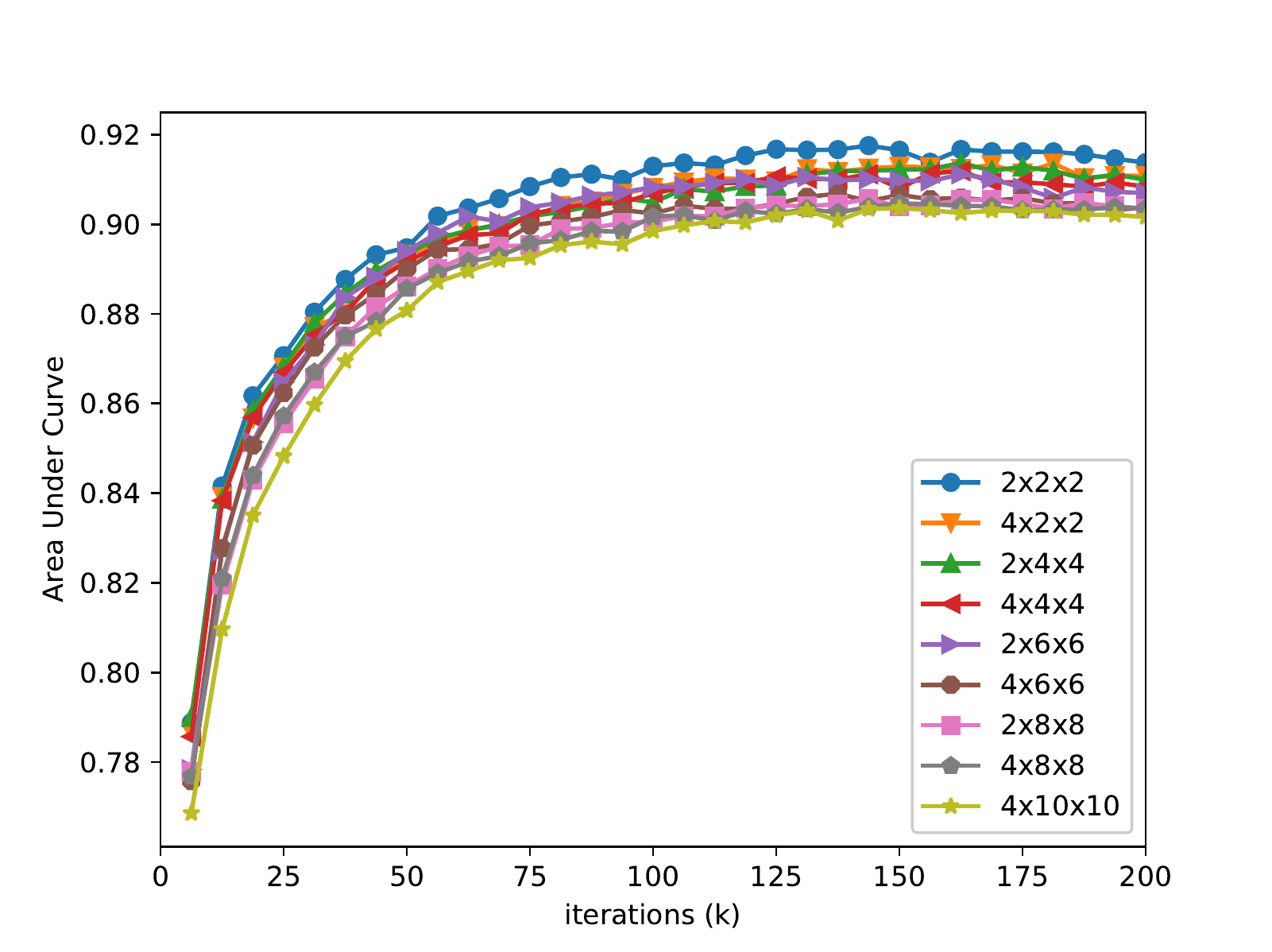}
	\caption{\label{fig:AUC for all granularities} The Area Under Curve (AUC) on the validation set vs. iteration steps on the training set for different granularities.}
\end{figure}

As is mentioned in Section \ref{section:Instrument conditions}, when deciding the granularity of voxels in the instrument, the optimal size may not be the smallest, and more factors should be taken into consideration. At the level of possible precisions achieved by our \emph{Topmetal} readout plane, transverse diffusion of the drifting charges and noise residing in the Charge Sensitive pre-Amplifier become constraints of choosing smaller size of voxels. So in practice, sometimes we use bigger size to trade off for better Signal Noise Ratio.

In the investigation of granularity effects, we only consider the ideal condition, with optimistic diffusion factors (1 mm for transverse, 2 mm for longitudinal) and no noise. The granularity generation mechanism is discussed in Section \ref{subsection:Data preprocessing}. We use the 2D\_base network to train and test on all the 9 granularities. Although we can directly work on 3D networks, we still choose to start with the 2D network, because 3D networks require intense computing power and huge storage space for small granularities, which makes them infeasible. On the other hand, we can use a single 2D network and some preprocessing methods to finish the granularity simulation. 

The mean value and the standard deviation of the accuracy and Area Under Curve (AUC) are shown in Fig. \ref{fig:Accuracy and AUC influenced by granularities}. AUC is a performance metric derived from the Receiver Operating Characteristic (ROC). Some examples of the ROC curves can be found in Fig. \ref{fig:ROC curve for different models}. We use AUC throughout the simulation studies because it represents the comprehensive capability of the classifier and shows stability at various iteration steps.

In Fig. \ref{fig:Accuracy and AUC influenced by granularities}, the y-axis values show that the impact of granularities on the accuracy and AUC is not significant. When using larger sizes, the accuracy tends to fall (within 2 percent), and the ROC curve (Fig. \ref{fig:ROC curve for group2x} and Fig. \ref{fig:ROC curve for group4x}) tends to degrade slightly, causing AUC to drop (within 0.015). The drop of AUC is consistent with the trend of increasing granularities, and it is much more stable than the accuracy metric.

The AUC performance on the validation set is shown in Fig. \ref{fig:AUC for all granularities}. In this figure, we can see different levels of discrimination capability achieved by various granularities. At nearly 100k iteration steps, AUCs of all granularities achieve their peak value, and the order of AUCs remains almost the same if we continue training. This shows that, increasing granularities can damage the discrimination power of the neural network model within an acceptable range. For grouped AUC curves, please refer to Fig. \ref{fig:AUC for group2x and group4x}.

\subsection{3D network simulation}
\label{subsection:3D network simulation}

\begin{table}[htbp]
	\centering
	\caption{Comparison of different networks and results in \cite{Renner:2016trj}. For a given percentage of signal events correctly classified, the table shows background events accepted as signals. Accuracy and AUC are also given.}
	\label{tab:Comparison of different networks}
	\begin{threeparttable}
		\small
		\begin{tabular}{lllll}
			\hline Analysis model (mm$^{3}$) & Iteration steps (k) & Accuracy (\%) & B.G. accepted (\%)\tnote{a} & AUC \\ 
			\hline 2D\_base (4$\times$10$\times$10) & 150 & 83.8 & 11.9 & 0.9036 \\
			       Conventional (10$\times$10$\times$5)\cite{Renner:2016trj} & --- & --- & 11.0 & --- \\
			       2D\_base (2$\times$2$\times$2) & 150 & 84.9 & 10.0 & 0.9165 \\
			       DNN (10$\times$10$\times$5)\cite{Renner:2016trj} & --- & --- & 9.4 & --- \\  
			       3D\_layer5 (8$\times$8$\times$8) & 125 & 86.5 & 8.6 & 0.9260 \\
			       3D\_layer7 (8$\times$8$\times$8) & 200 & 86.8 & 6.6 & 0.9381 \\
			       3D\_Res (8$\times$8$\times$8) & 50 & 91.1 & 4.1 & 0.9590 \\
			\hline 
		\end{tabular} 
		
		\begin{tablenotes}
			\footnotesize
			\item[a] Calculated at 76.6\% Signal efficiency.		
		\end{tablenotes}	
	\end{threeparttable}
\end{table}

\begin{figure}[htbp]
	\centering
	\subfigure[full]{
		\includegraphics[width=0.48\textwidth]{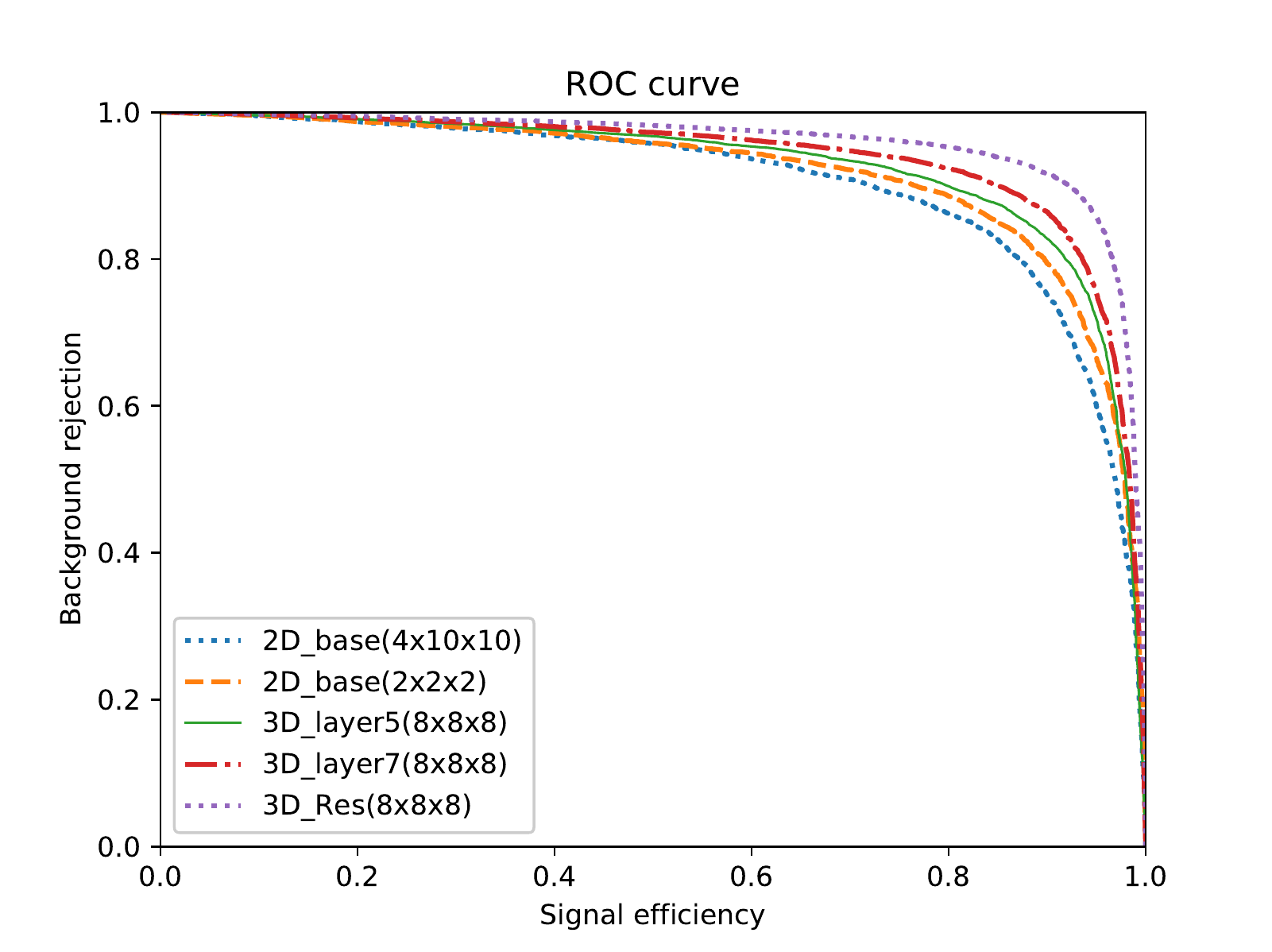}}	
	\subfigure[high B.G.]{
		\includegraphics[width=0.48\textwidth]{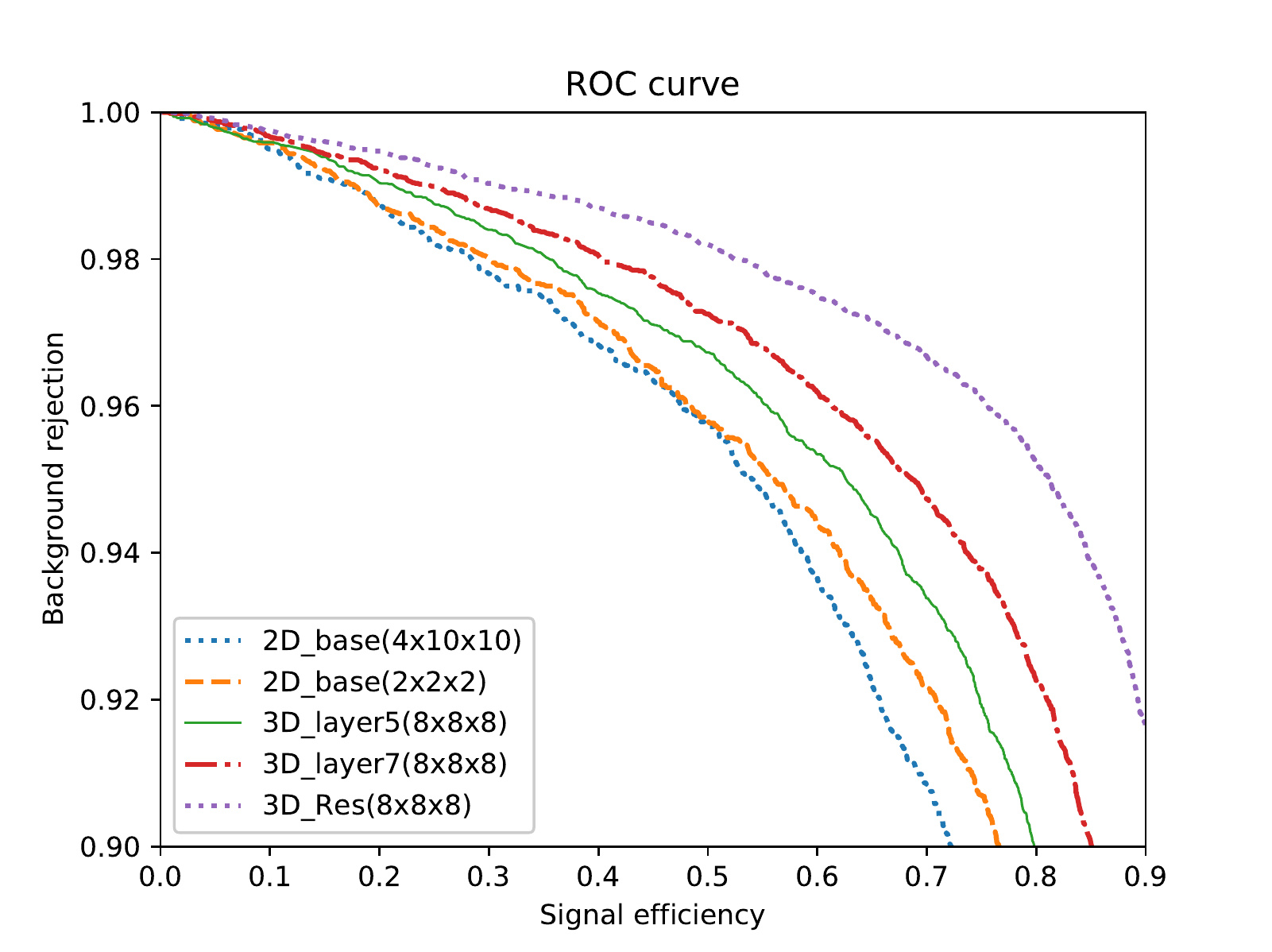}}
	\caption{ROC curves for different models. (a) The full ROC curves. (b) The ROC curves at high background rejection.}
	\label{fig:ROC curve for different models}
\end{figure}

To test our 3D CNNs on the 3D cuboid dataset, we do the 3D network simulations and compare to the previous 2D network and results in a recent work \cite{Renner:2016trj}. The 3D CNNs we use are discussed in Section \ref{section:Network architectures} and the data preprocessing is described in Section \ref{subsection:Data preprocessing}. The granularity is fixed to 8 mm $\times$ 8 mm $\times$ 8 mm for all 3D networks. This voxel size can be achieved by our \emph{Topmetal} readout plane by a large margin and is currently estimated to be the optimal size for constructing detection instruments and for data analysis.

The main result is shown in Table \ref{tab:Comparison of different networks} and Fig. \ref{fig:ROC curve for different models}. In the table, a total of 7 analysis models are involved. "Conventional" refers to a method which used a man-made energy threshold to judge whether one or two "blobs" of energy are presented in the reconstructed track. Usually $0\nu\beta\beta$ events have two ends with energy depositions, while $beta$ events have only one end with obvious energy depositions. DNN refers to GoogLeNet \cite{szegedy2015going}, an intricate 2D convolutional neural network with "inception" structures and 22 parameterized layers. It has about $7 \times 10^{6}$ parameters. In comparison, our 3D CNNs (3D\_layer5, 3D\_layer7 and 3D\_Res) have about $1 \times 10^{6}$ parameters each, which is only 1/7 of the GoogLeNet. The granularity for the conventional method and DNN is 10 mm $\times$ 10 mm $\times$ 5 mm, and it has a volume of 500 $mm^{3}$, comparable to 512 $mm^{3}$ in the case of 8 mm $\times$ 8 mm $\times$ 8 mm granularity.

In the table, the signal efficiency is fixed to a given value to calculate the ratio of background events being accepted. We can see a significant gain resulting from both the 3D architecture and the depth of the neural network. When we use the plain 2D CNN, the result at the minimum granularity (10.0\%) is slightly worse than the DNN (9.4\%). However, when we use 3D\_layer5 with the similar architecture, the background acceptance rate can easily decrease to 8.6\%, which is below DNN. Furthermore, when we add more layers to the model, the results can be even better. The best result we achieve currently is 4.1\% under 3D\_Res. At high signal efficiency rates, a several percentage increase can lead to significantly better ROC curves. In the figure which corresponds to the table, the curves of different models are well distinguished, and the improvement of AUC can be seen clearly. In the best case of 3D\_Res, when the signal efficiency is set to 90\%, the background rejection can still be above 90\% (below 10\% background events are accepted). This is the desired effect of a classifier used to reject background events when signal events are scarce.

\subsection{Diffusion and noise simulation}
\label{subsection:Diffusion and noise simulation}

\begin{table}[htbp]
	\centering
	\caption{Test results for different diffusion and noise conditions.}
	\label{tab:Diffusion and noise}
	\scriptsize
	\begin{tabular}{llllll}
		\hline  Diffusion factor (mm) & & Noise factor ($e^-$) & Accuracy (\%) & Area Under Curve & F1 score (\%) \\
		\cline{1-2} Transverse & Longitudinal & & & & \\
		\hline  1.0 & 2.0 & 0.0 & \textbf{90.4$\pm$0.9} & \textbf{0.9577$\pm$0.0017} & \textbf{90.4$\pm$1.2} \\ 
		1.0 & 2.0 & 10.0 & 89.8$\pm$1.5 & 0.9541$\pm$0.0020 & 89.6$\pm$2.0 \\ 
		1.0 & 2.0 & 30.0 & 89.0$\pm$2.1 & 0.9512$\pm$0.0024 & 88.7$\pm$2.8 \\ 
		\hline  1.0 & 1.0 & 0.0 & 90.0$\pm$1.3 & 0.9567$\pm$0.0005 & 90.3$\pm$1.0 \\ 
		1.0 & 2.0 & 0.0 & \textbf{90.4$\pm$0.9} & \textbf{0.9577$\pm$0.0017} & \textbf{90.4$\pm$1.2} \\
		1.0 & 5.0 & 0.0 & 89.8$\pm$1.9 & 0.9557$\pm$0.0015 & 90.2$\pm$1.4 \\ 
		1.0 & 9.0 & 0.0 & 89.8$\pm$1.6 & 0.9556$\pm$0.0011 & 89.7$\pm$2.1 \\ 
		\hline  9.0 & 1.0 & 0.0 & 90.4$\pm$0.3 & 0.9551$\pm$0.0010 & 90.5$\pm$0.5 \\ 
		9.0 & 2.0 & 0.0 & \textbf{90.5$\pm$0.2} & \textbf{0.9569$\pm$0.0010} & \textbf{90.7$\pm$0.1} \\ 
		9.0 & 5.0 & 0.0 & \textbf{90.5$\pm$0.2} & 0.9550$\pm$0.0007 & 90.6$\pm$0.4 \\ 
		9.0 & 9.0 & 0.0 & 90.1$\pm$0.2 & 0.9517$\pm$0.0009 & 90.3$\pm$0.1 \\ 
		\hline  9.0 & 9.0 & 30.0 & 89.7$\pm$0.2 & 0.9488$\pm$0.0014 & 89.8$\pm$0.2 \\
		\hline 
	\end{tabular}
\end{table}

In this part, we use datasets in more realistic situations; the transverse and longitudinal diffusion of the drifting charge cloud is taken into consideration, as well as the noise of the Charge Sensitive pre-Amplifier. Both transverse and longitudinal diffusion is significant in pure xenon, which may damage the spatial resolution achieved by the readout system. Furthermore, random noise in the detector can result in non-zero values (which could be negative), which poses a challenge for triggering and preprocessing.

The network architecture we use in this part is 3D\_Res, and the result is shown in Table \ref{tab:Diffusion and noise}. The diffusion and noise factors are similar to those expected in the TPC discussed in Section \ref{section:Instrument conditions}. The diffusion factors represent the standard deviations of the diffusion, and the noise factors represent the noise level collected along the z axis. Noticing that the noise values of voxels obey the Gaussian distribution and are independent of each other, we estimate the sigma of Gaussian noise collected along the z axis, which can be denoted as $S_{all}$. Then, the sigma for each voxel can be calculated by $S_{voxel}=\frac{S_{all}}{\sqrt{N_{z}}}$, in which $N_{z}$ is the total number of voxels along the z axis. F1 score in the table is a comprehensive metric of the capability of the classifier. First, we compute $Precision=\frac{TP}{TP+FP}$ and $Recall=\frac{TP}{TP+FN}$, in which TP stands for the number of "true" positive/signal records, FP stands for the number of "false" positive/signal records and FN stands for the number of "false" negative/background records. Then the F1 score is calculated by $\frac{2*Precision*Recall}{Precision+Recall}$.

For each row in the table, we generate the dataset, preprocess it, train the neural network and test performance metrics in a separate procedure. In order to get comparable results for each variable, the rows in the table within two horizontal lines use the same parameter when preprocessing. In general, diffusion and noise don't cause obvious changes in the performance of the 3D CNN. We can see the difference between several conditions is very slight, and randomness of the training process may contribute to the accuracies. Nonetheless, we can still make the following arguments:

\begin{enumerate}
	
	\item Noise is not favorable for the neural network model, but the impact is acceptable within the estimated range of the detector.
	
	\item When transverse diffusion is fixed, the best results for different longitudinal diffusions don't appear at the least significant value. We can see a rise at the beginning, and a fall after crossing a threshold. In other words, some amount of diffusion seems to be necessary to reach the optimal performance by the neural network model.
	\label{item:rise and fall}
	
	\item Compared to the baseline condition (the last row in the table), the influence of the diffusion and noise is limited in one or two percent.
	
\end{enumerate}

For \ref{item:rise and fall}, it can be seen that although low diffusion factors can preserve more physical information, they may not be the best for pattern recognition. If the charge distribution in the dataset is too concentrated (the first row in the second section) or too discrepant (the fourth row in the second section and the first row in the third section), the overall classification ability is impacted by the "extreme" cases. In fact, the fluctuation of the group with 1 mm transverse factor is significantly larger than the group with 9 mm transverse factor, which indicates that too intense features in the input cuboid will damage the stability of the neural network model. The optimal result is achieved by trading off between the integrity of physical information and the legibility of the input pattern. Please refer to Fig. \ref{fig:AUC for noise, t1 and t9} for graphs demonstrating the performance and fluctuation at different diffusion and noise levels.

\section{Discussion and conclusion}
\label{section:Discussion and conclusion}

Machine learning techniques are regarded as an important supplement to traditional physical analysis in high energy physics. Their applications include collision reconstruction \cite{atlas2014neural}, event interpretation \cite{almeida2015playing} and low-level triggering probably in the future. The most common used methods are boosted decision trees \cite{aaij2013lhcb,gligorov2013efficient} and neural networks \cite{baldi2014searching}. When dealing with a large amount of data, machine learning techniques are expected to approximate or even exceed human performance.

In this study, a machine learning approach validated by previous image classification tasks is applied to solve a particular physical problem: how to improve the instrument efficiency in the search for neutrinoless double-beta decay and reject background events. Specifically, we adopt a deep learning method, and the architectures we use are 3D convolutional neural networks and 3D residual neural networks. The simulation results are competitive records in the field. There is an optimal value of the voxel size which is detector dependent. For our simulation, it is 8 mm $\times$ 8 mm $\times$ 8 mm, because it fits the experiment requirements and makes a good trade-off between the detector noise and the discrimination ability. In contrast to collision experiments where events can be generated arbitrarily given proper experimental condition and enough running time, double-beta decay events are so rare that we need a considerable amount of expensive isotope material and years of observation. A good classifier can effectively reduce the time needed to form convincing conclusions and the total cost of maintaining the instrument. It is of particular importance in the case where both time and resources are limited.

Unlike decision trees and many other traditional machine learning techniques, neural networks are not so interpretable, especially when dealing with a complex question when many factors are variable, such as dataset probability distributions, preprocessing parameters, network architectures and so on. Moreover, well-designed models can only recognize some patterns with best discrimination power. However, these patterns may not preserve as much physical information as possible. How to disentangle these factors and achieve the optimal settings could involve much more computation and remain a meaningful piece of work in the future.

\appendix

\section{Details of the toy model}
\label{section:Details of the toy model}

\subsection{The probabilistic model}

We treat the event in the Time Projection Chamber as a segmented "chain". The passage of the high-energy electron is divided into some small parts, each of which is considered to be an ideal line segment. For signal events ($0\nu\beta\beta$), two electron passages are connected at a common joint and each of them is terminated with obvious energy depositions (Bragg Peak). For background events ($beta$), no joint is presented in the chain and there are obvious energy depositions at only one end of the chain. Besides, we also consider partial models, with only segments or only energy depositions. The partial models are mainly used for the factor decomposition analysis.

The total energy for signal and background events is set to be identical. For a signal event, the energy is carried away by the two electrons; for a background event, the single electron carries all the energy.

To reduce the complexity of the toy model, we make following assumptions:

\begin{enumerate}

	\item (Naive Bayes assumption) Generally, the scattering angles and the energy depositions are correlated by the energy of the electron. Here for simplicity, we assume that the scattering angles and the energy depositions both obey the Gaussian distribution, with parameters mutually independent of each other.
	\item (Sum of energy assumption) In the original form of the toy model, the two deposition ends of the signal event and the single deposition end of the background event are represented by separate random variables. To make the probabilistic model compatible with both signal events and background events, we assume that the energy of depositions sum up to a fixed value. Thus, we only need one variable to represent the depositions of the signal event.
	\item (2D assumption) Here we only want to demonstrate the discrimination ability of deep learning methods, it is convenient to use 2D models to facilitate the synthetic data generation.

\end{enumerate}

Based on the assumptions above, the generative model for background events and signal events can be written as follows:

\begin{equation}
	\textbf{x} = (x_1, x_2, ..., x_i, ..., x_{N-1})
\end{equation}

\begin{equation}
	f(\textbf{x}, y, N, V=b) = P(V=b) P(N) \left(\prod_{i=1}^{N-1}{f_{X}(x_i|0, \sigma_{bt})}\right) g_{Y}(y|\mu_{bd}, \sigma_{bd}) \\
\end{equation}

\begin{equation}
\begin{split}
	f(\textbf{x}, y, N, N_1, V=s) = &P(V=s) P(N, N_1) \left(\prod_{i=1}^{N_1-1}{f_{X}(x_i|0, \sigma_{st})}\right) f_{X^{\prime}}(x_{N_1}|0, \sigma_{si}) \\
		                           &\left(\prod_{i=N_1+1}^{N-1}{f_{X}(x_i|0, \sigma_{st})}\right) g_{Y}(y|\mu_{sd}, \sigma_{sd})
\end{split}
\end{equation}

\noindent where $\textbf{x}$ denotes angles in the segmented chain, $y$ denotes energy depositions at the far end, $N$ denotes total number of segments, $N_1$ denotes segments before the joint for signal events, and $V$ denotes the event type ($b$ for background and $s$ for signal). $f_X$, $f_{X^{\prime}}$ and $g_Y$ are probability density functions of Gaussian distributions, conditioned on the mean value and the standard deviation.

Parameters for different numbers of segments are listed in Table \ref{tab:Paramters for segments}. $\mu$ represents the mean value and $\sigma$ represents the standard deviation. The unit of angles is radian, and the depositions are expressed as the ratio of total energy. The values are estimated from the physical process with some relaxation \cite{bethe1953moliere}. Since multiple scattering is significant for $\sim$ MeV electrons in 10 bar xenon gas, it is impractical to fit the angles into the toy model directly. Instead, we use a small-angle estimation which is rather conservative, and this is possibly a limitation of our model.

\begin{table}[htbp]
	\centering
	\caption{Parameters for different numbers of segments.}
	\label{tab:Paramters for segments}
	\begin{tabular}{llllllll}
		\hline Segments & $\sigma_{bt}$ & $\mu_{bd}$ & $\sigma_{bd}$ & $\sigma_{st}$ & $\sigma_{si}$ & $\mu_{sd}$ & $\sigma_{sd}$ \\ 
		\hline N=2 & 0.31 & 0.80 & 0.07 & 0.61 & 0.79 & 0.50 & 0.17 \\
		       N=3 & 0.25 & 0.80 & 0.07 & 0.49 & 0.79 & 0.50 & 0.17 \\
		\hline 
	\end{tabular} 
\end{table}

To calculate the theoretical upper limit for the complete model with both segments and depositions, the \emph{a posteriori} probability is given by:

\begin{equation} \label{equ:Posteriori}
	P(V=v_i|\textbf{x}, y, N) = \frac{f(\textbf{x}, y, N, V=v_i)}{f(\textbf{x}, y, N)} \sim f(\textbf{x}, y, N, V=v_i)
\end{equation}

\noindent where $N_1$ is marginalized, and the normalizer $f(\textbf{x}, y, N)$ is the same for both events. Finally, the theoretical limit is given by:

\begin{equation} \label{equ:Complete model}
	P^{*}_{com} = \sum_{i=1}^{2}{\idotsint_{C_i}{f(\textbf{x}, y, V=v_i|N) \mathrm{d}x_1\mathrm{d}x_2\cdots\mathrm{d}x_{N-1}\mathrm{d}y}}
\end{equation}

\begin{equation} \label{equ:Mixing model}
	P^{*}_{mix} = \sum_{i=1}^{2}\sum_{N}{\idotsint_{C_i}{f(\textbf{x}, y, N, V=v_i) \mathrm{d}x_1\mathrm{d}x_2\cdots\mathrm{d}x_{N-1}\mathrm{d}y}}
\end{equation}

\noindent where $C_i$ denotes the region where the class i has the highest posterior. Equation \ref{equ:Complete model} is used for complete models when $N$ is fixed (the fourth row and the fifth row in Table \ref{tab:Toy model and limits}), and Equation \ref{equ:Mixing model} is used for the mixing model when $N$ varies (the sixth row in Table \ref{tab:Toy model and limits}). To calculate the theoretical limit for partial models (the first three rows in Table \ref{tab:Toy model and limits}), the method is almost the same. We marginalize $y$ or $\textbf{x}$ before calculating the \emph{a posteriori} probability (Equation \ref{equ:Posteriori}) and the theoretical limit (Equation \ref{equ:Complete model}).

\subsection{Data generation for the CNN}

When casting the toy model to 2D images, we generate signal and background events on an 80 $\times$ 80 canvas. The segments are represented by lines with a certain width, and the deposition ends are represented by filled circles with a certain radius. The intensity of lines and circles is properly chosen to reflect the different rates of energy loss. The angles and energy depositions are randomly sampled from the aforementioned Gaussian distributions. For partial models, either circles or lines are omitted when we draw on the canvas.

The original granularity for both events is 2 mm $\times$ 2 mm. Then we blur the image and reduce its granularity to 8 mm $\times$ 8 mm using data preprocessing techniques discussed in Section \ref{subsection:Data preprocessing}. We also tried more fine-grained input images but didn't observe significant improvements in the performance. After that, the 2D images are fed to the 2D\_toy neural network for training and testing. The training dataset has 40000 samples and the test dataset has 10000 samples. Other configuration issues are similar to what is discussed in Appendix \ref{section:Configuration and implementation}.

\section{Configuration and implementation}
\label{section:Configuration and implementation}

For the 2D CNN and the granularity simulation, we use 100,000 simulation samples as our training dataset and 12,000 samples as our validation dataset. For 3D CNNs and subsequent simulations, we use 100,000 simulation samples as our training dataset and 20,000 samples as our validation dataset. The signal events and background events in the simulation dataset are balanced, i.e., each category takes up 50\% of the training and validation dataset. The optimization algorithm we use is stochastic gradient descent with initial learning rate 0.001, and the minibatch we use is 32. We train the networks from scratch without any pre-trained models. Weights in the filter and the fully-connected layer are initialized to zero mean, 0.005$\sim$0.01 standard deviation Gaussian distribution, while biases are initialized to zero or a small constant (0.1). We pause training at intervals of a whole number of epoches, and test our model on the validation dataset. Dropout \cite{srivastava2014dropout} is used to overcome the overfitting problem. All the neural networks are implemented using TensorFlow \cite{abadi2016tensorflow}, an open-source software framework designed and maintained by Google, Inc. To speed up training, simulations were conducted on a desktop with Intel(R) Core(TM) i7-5820K CPU @ 3.60 GHz, 64 GB DDR4 RAM, and a single NVIDIA GeForce GTX TITAN X GPU with 12GB video memory.

\section{Supplements to the simulation studies}

\subsection{Supplements to the granularity simulation}
\label{subsection:Supplements to the granularity simulation}

\begin{figure}[H]
	\centering
	\subfigure[2 mm for x]{
		\includegraphics[width=0.48\textwidth]{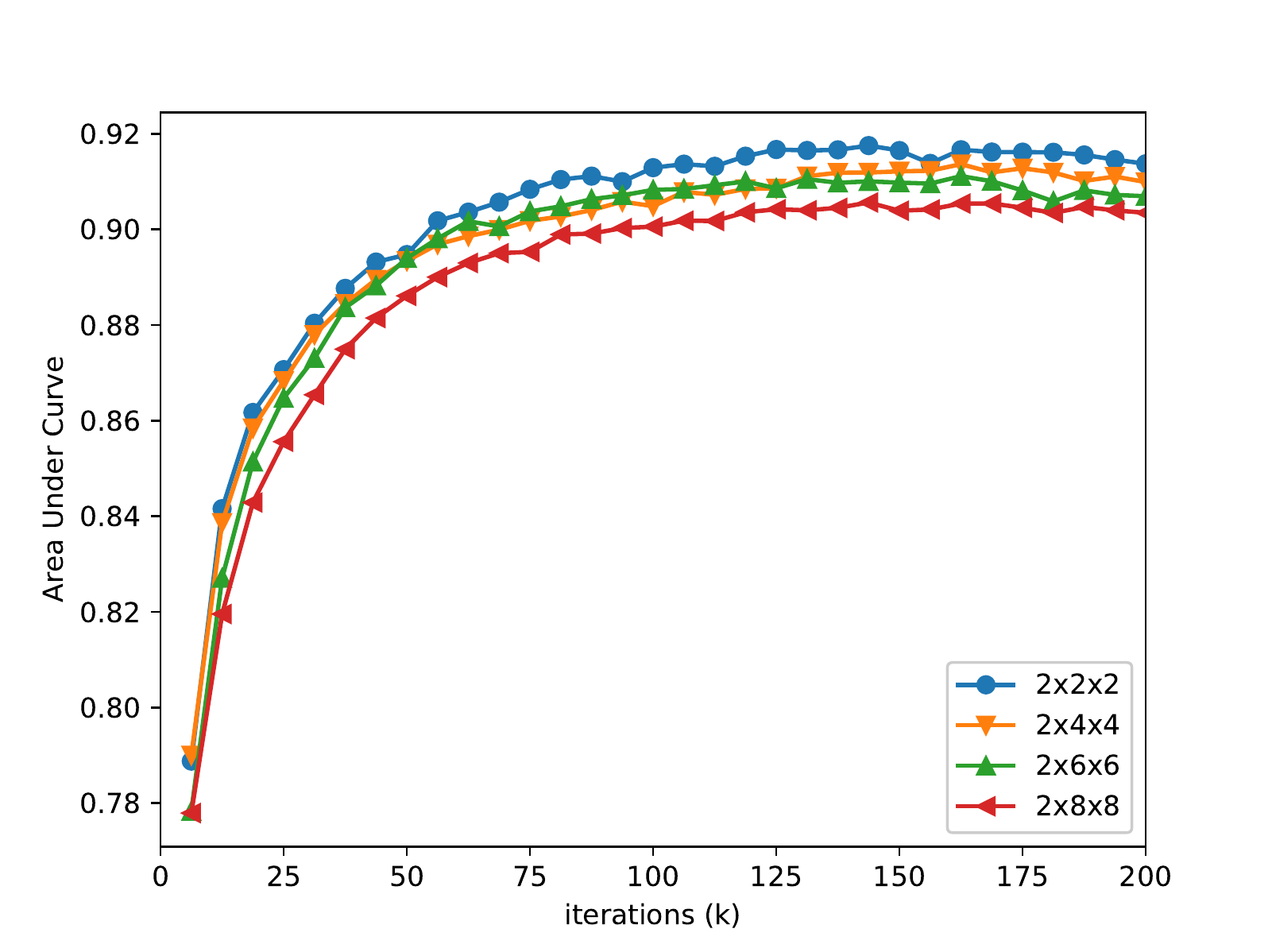}}
	\subfigure[4 mm for x]{
		\includegraphics[width=0.48\textwidth]{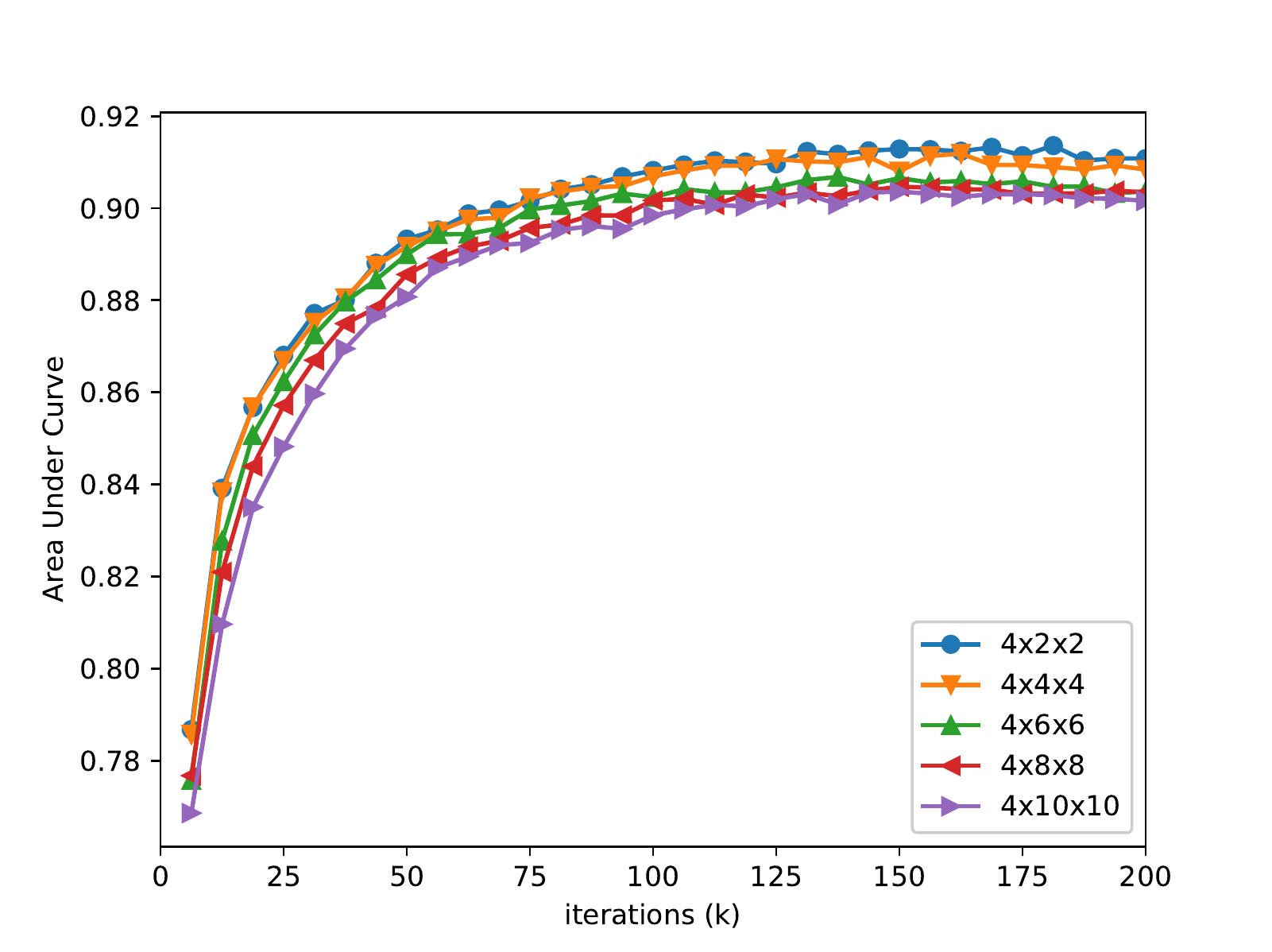}}
	\caption{\label{fig:AUC for group2x and group4x} AUC on the validation set vs. iterations on the training set for different granularities. In these figures, we can see that the AUC performance reaches a plateau at about 100k iteration steps, and AUC is consistent with the size of granularities.}
\end{figure}

\begin{figure}[H]
	\centering
	\subfigure[2 mm, full]{
		\includegraphics[width=0.48\textwidth]{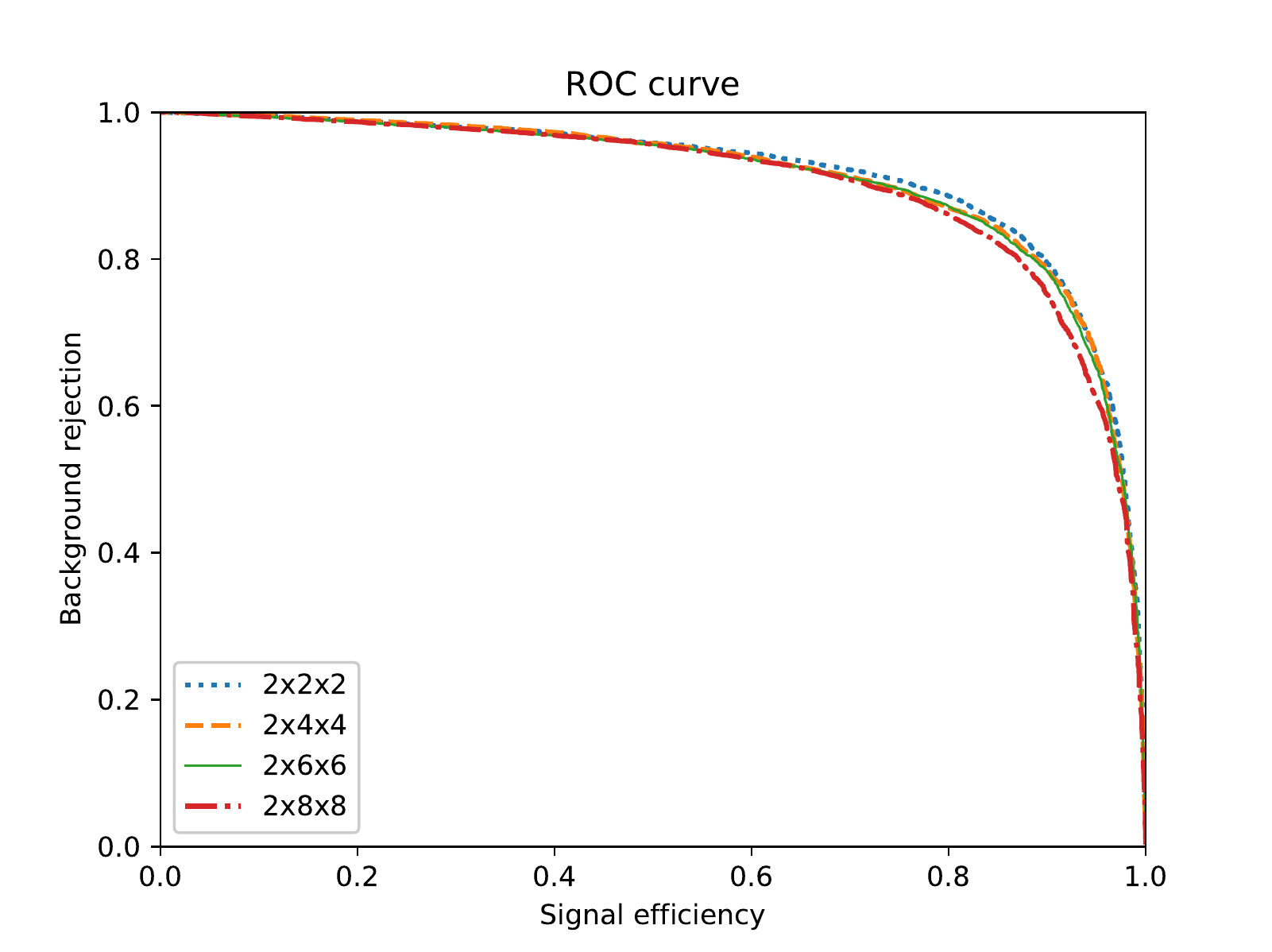}}
	\subfigure[2 mm, high B.G.]{
		\includegraphics[width=0.48\textwidth]{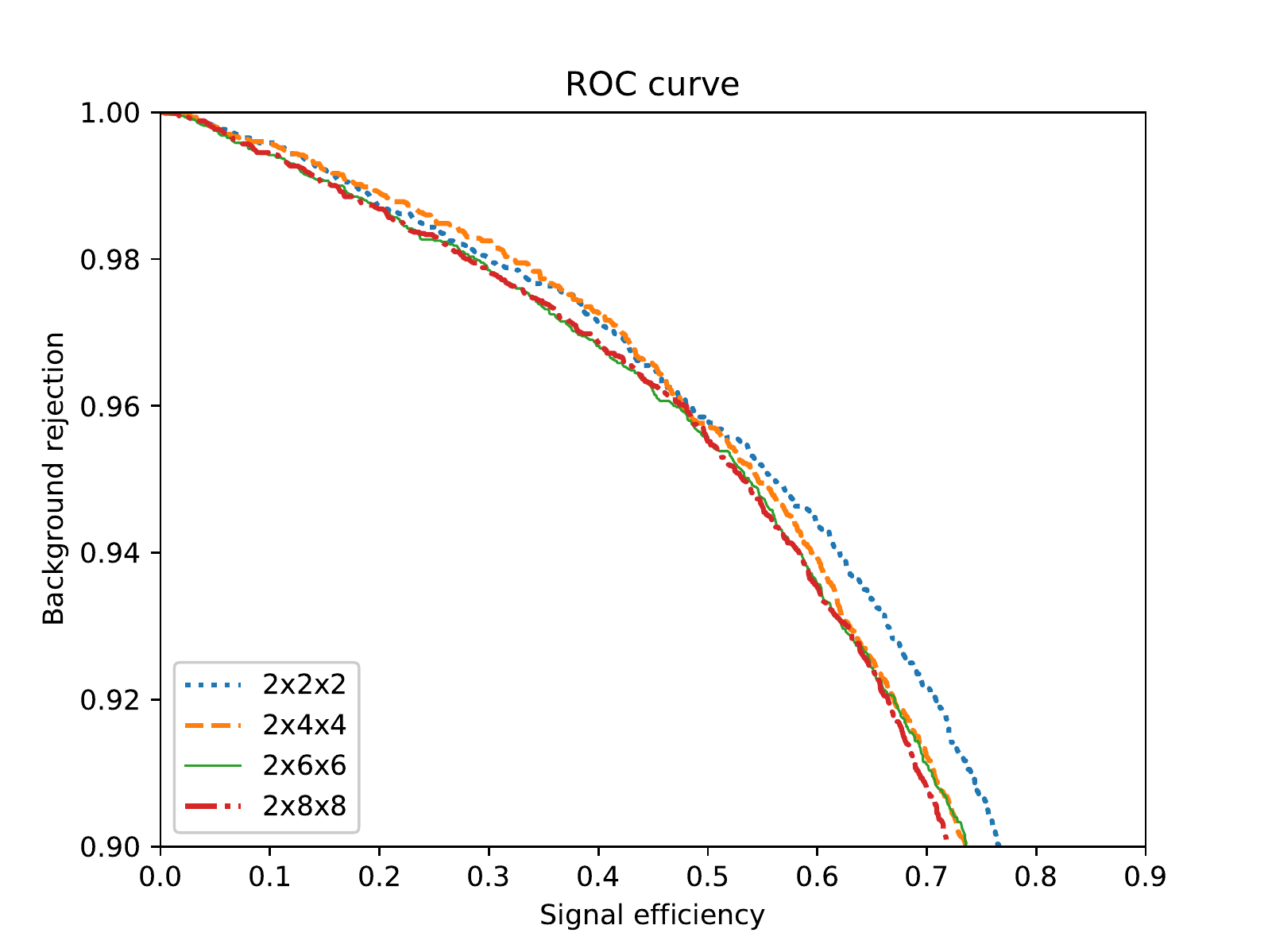}}
	\caption{\label{fig:ROC curve for group2x} The full ROC curves and the ROC curves at high background rejection rate for 2 mm size along x axis at 150k iteration steps. In the left figure, ROC curves with different granularities are slightly distinguishable and show accordance with the size of granularities. In the right figure, the ROC curves are chaotic at extremely high background rejection rate. However, starting from a point ($\sim$0.96 background rejection), the influence of granularities seems to dominate the order of ROC curves.}
\end{figure}

\begin{figure}[H]
	\centering
	\subfigure[4 mm, full]{
		\includegraphics[width=0.48\textwidth]{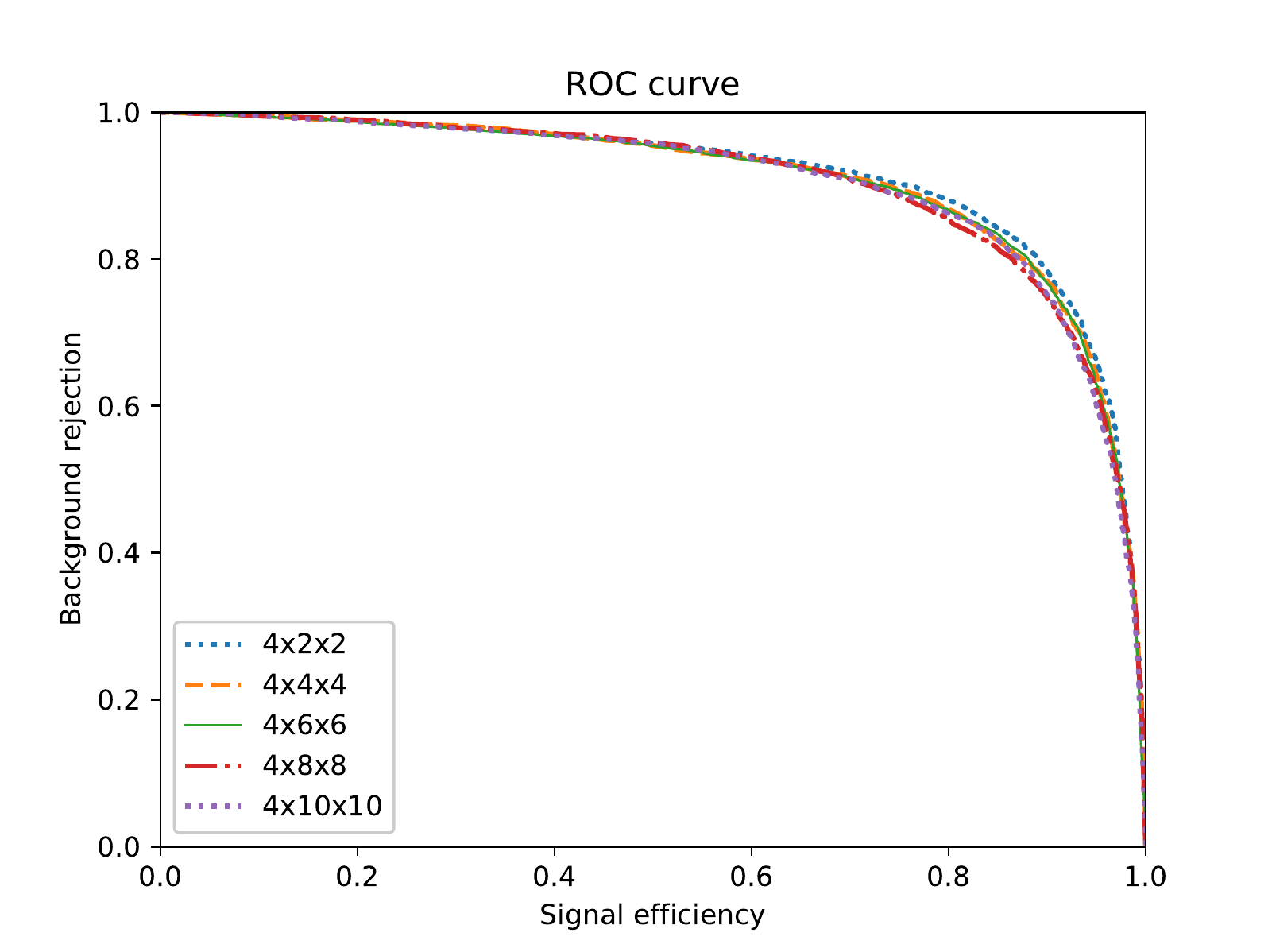}}
	\subfigure[4 mm, high B.G.]{
		\includegraphics[width=0.48\textwidth]{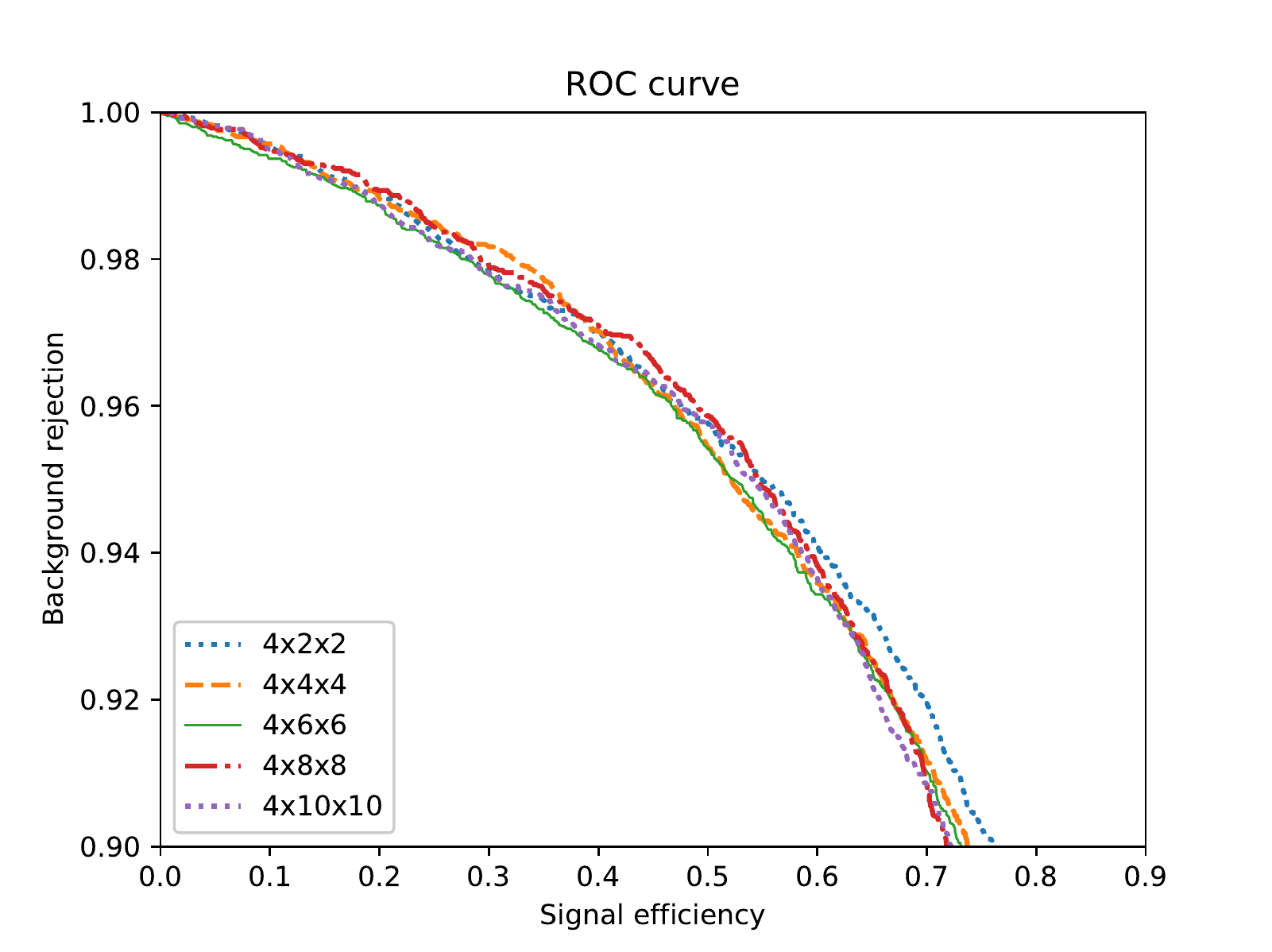}}
	\caption{\label{fig:ROC curve for group4x} The full ROC curves and the ROC curves at high background rejection rate for 4 mm size along x axis at 150k iteration steps. In the left figure, ROC curves with different granularities are slightly distinguishable and show accordance with the size of granularities in most regions. In the right figure, the chaotic region is larger than Fig. \ref{fig:ROC curve for group2x}. Starting from a point ($\sim$0.92 background rejection), the influence of granularities seems to dominate the order of ROC curves.}
\end{figure}

\subsection{Supplements to the diffusion and noise simulation}
\label{subsection:Supplements to the diffusion and noise simulation}

\begin{figure}[H]
	\centering
	\subfigure[AUC for noise]{
		\includegraphics[width=0.48\textwidth]{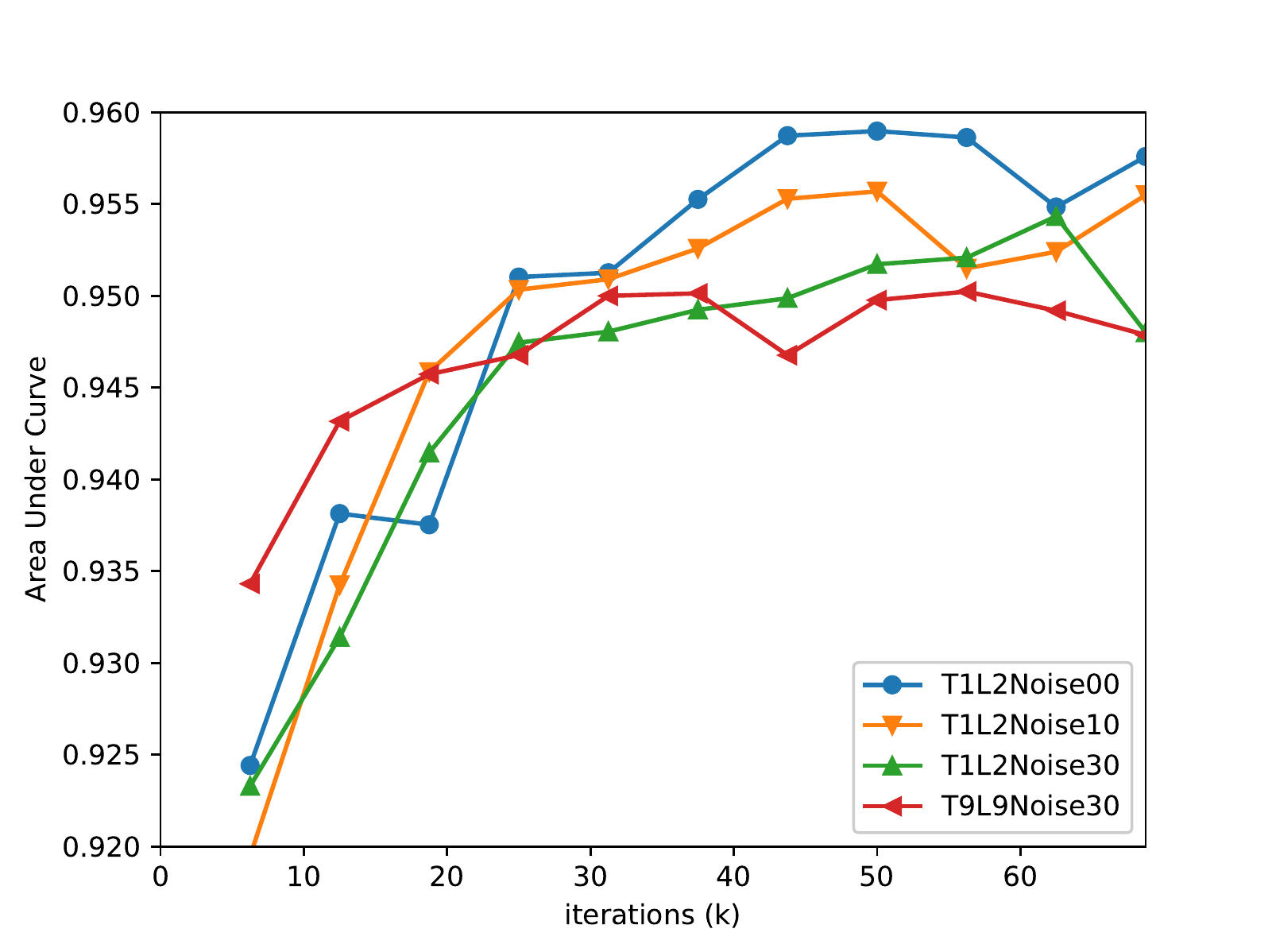}}
	\subfigure[AUC for 1 mm transvere diffusion factor]{
		\includegraphics[width=0.48\textwidth]{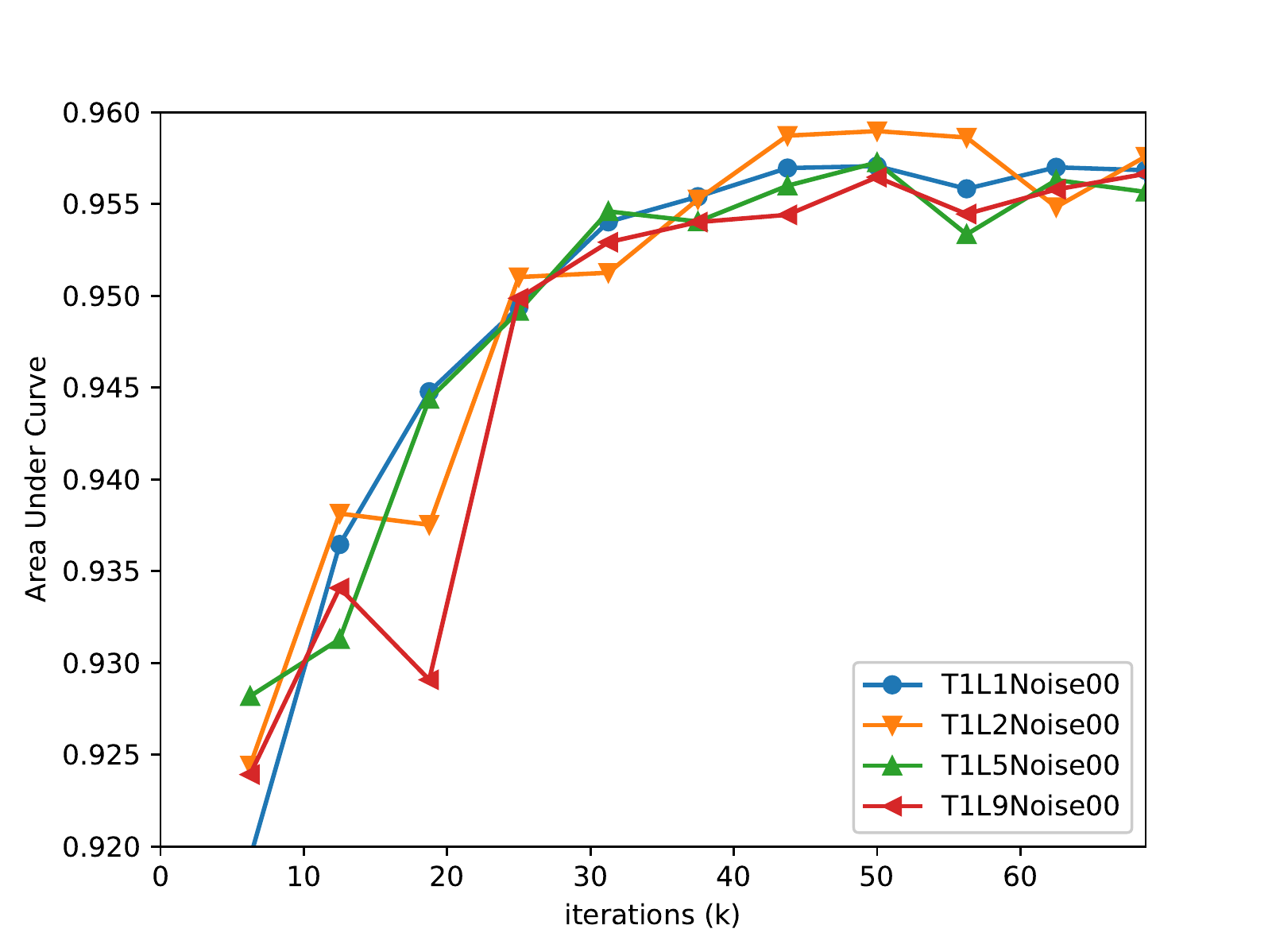}}
	\subfigure[AUC for 9 mm transvere diffusion factor]{
		\includegraphics[width=0.48\textwidth]{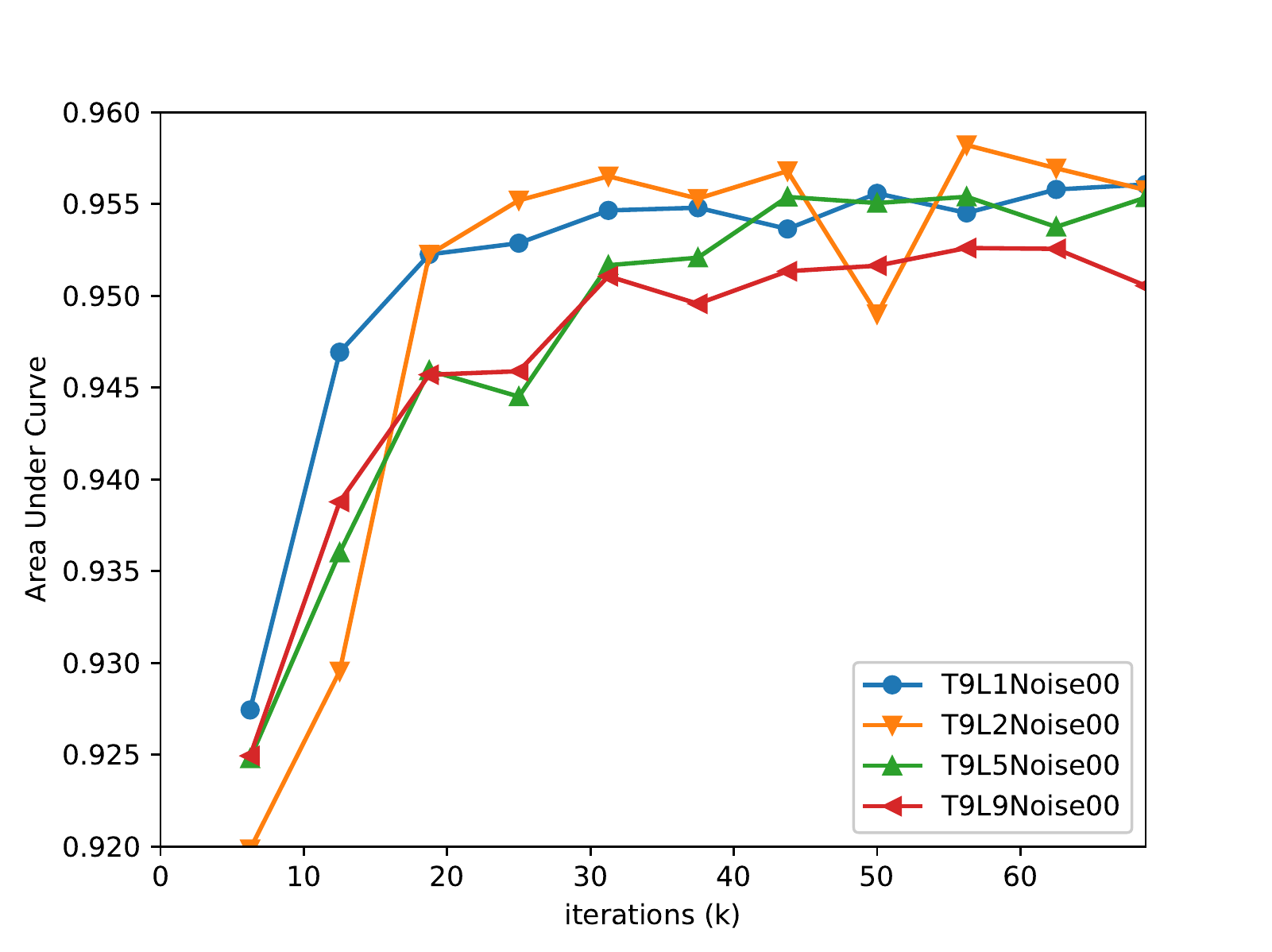}}
	\caption{\label{fig:AUC for noise, t1 and t9} AUC on the validation set vs. iterations on the training set for different noise levels and diffusion factors. These curves show larger fluctuations than curves in Fig. \ref{fig:AUC for group2x and group4x}.}
\end{figure}

\acknowledgments

This research is supported by the National Natural Science Foundation of China (Grant Number 11505074). We thank Yuan Mei from Lawrence Berkeley National Laboratory for providing the source code to generate the simulation data and giving valuable comments on the manuscript.


\providecommand{\href}[2]{#2}\begingroup\raggedright\endgroup

\end{document}